\documentclass[useAMS,usenatbib]{mn2e}

\newcommand{\phref}[3]{\textcolor{#1}{\href{#2}{#3}}}

\usepackage[dvips]{graphicx}
\usepackage{epsfig}
\usepackage{txfonts}
\usepackage{color}
\usepackage{setspace}
\usepackage{multirow}
\usepackage{hyperref}

\newcommand{\mygraphics}[2]{\includegraphics[#1]{PDF_Figures/#2}}

\newcommand{\dg}{$^{\circ}$}

\newcommand{\pd}[2]{\frac{\partial #1}{\partial #2}}
\newcommand{\Pdiv}[1]{{\vec{\nabla}}\cdot (\,#1\,)}
\newcommand{\Pgrad}[1]{{\vec{\nabla}}#1}
\newcommand{\pvec}[1]{{\vec{#1}}}
\newcommand{\rmnum}[1]{\romannumeral #1}

\newcommand{\cygx}[0]{Cyg\,\,X-3}
\newcommand{\J}[1]{{\em Jet\,#1}}
\newcommand{\SNR}[1]{{\em SNR\,#1}}
\newcommand{\SJ}[1]{{\em SNRJet\,#1}}
\newcommand{\zd}[0]{Z_{\rm d}}

\newcommand{\tlab}[1]{\label{tab:#1}}
\newcommand{\flab}[1]{\label{fig:#1}}
\newcommand{\elab}[1]{\label{eq:#1}}

\newcommand{\slab}[1]{\label{sec:#1}}

\newcommand{\tref}[1]{\ref{tab:#1}}
\newcommand{\fref}[1]{Figure \ref{fig:#1}}
\newcommand{\eref}[1]{(\ref{eq:#1})}
\newcommand{\cref}[1]{chapter \ref{chap:#1}}
\newcommand{\sref}[1]{\S \ref{sec:#1}}

\newcommand{\corx}[1]{#1}

\title[Hydrodynamic Simulations of the SS\,433-W\,50 Interaction]{When Microquasar Jets and Supernova Collide:\\ Hydrodynamically  Simulating the SS\,433-W\,50 Interaction}
\author[P.\,T. Goodall, F. Alouani-Bibi and K. Blundell]{Paul T. Goodall$^{1}$\thanks{E-mail:
ptg@astro.ox.ac.uk (PTG)}, Fathallah Alouani-Bibi$^{2}$\thanks{E-mail: alouani@physics.gmu.edu (FAB)} and Katherine M.\ Blundell$^{1}$\thanks{E-mail: kmb@astro.ox.ac.uk (KMB)}\\
$^{1}$Astrophysics, University of Oxford, Keble Road, Oxford, OX1 3RH.\\
$^{2}$Department of Physics and Astronomy, George Mason University,  4400 University Drive, Fairfax VA, 22030 USA.}

\begin{document}
\date{Accepted 1988 December 15. Received 1988 December 14; in original form 1988 October 11}


\maketitle
\label{firstpage}

\begin{abstract}
We present investigations of the interaction between the relativistic, precessing jets of the microquasar SS\,433 with the surrounding, expanding Supernova Remnant (SNR) shell W\,50, and the consequent evolution in the inhomogeneous Interstellar Medium (ISM).   We model their evolution using the  hydrodynamic FLASH code, which uses adaptive mesh refinement.  We show that the peculiar morphology of the entire nebula can be reproduced to a good approximation, due to the combined effects of: (i) the evolution of the SNR shell from the free-expansion phase through the Sedov blast wave in an exponential density profile from the Milky Way disc, and (ii) the subsequent interaction of the relativistic, precessing jets of SS\,433.   Our simulations reveal:  (1) Independent measurement of the Galaxy scale-height and density local to SS\,433 (as $n_{\rm 0}=0.2\,{\rm cm}^{-3},\zd=40\,{\rm pc})$, with this scale-height being in excellent agreement with the work of Dehnen \& Binney.  (2)  A new mechanism for hydrodynamic refocusing of conical jets.  (3) The current jet precession characteristics do not simply extrapolate back to produce the lobes of W\,50 but a history of episodic jet activity having at least 3 different outbursts with different precession characteristics would be sufficient to produce the W\,50 nebula.   A history of intermittent episodes of jet activity from SS\,433 is also suggested in a kinematic study of W\,50 detailed in a companion paper (Goodall et al, MNRAS submitted). (4) An estimate of the age of W\,50, and equivalently the age of SS\,433's black hole created during the supernova explosion, in the range of 17,000 - 21,000 years.  
\end{abstract}



\begin{figure*}
\centering
\begin{minipage}{\textwidth}
\centering
\mygraphics{width=\textwidth}{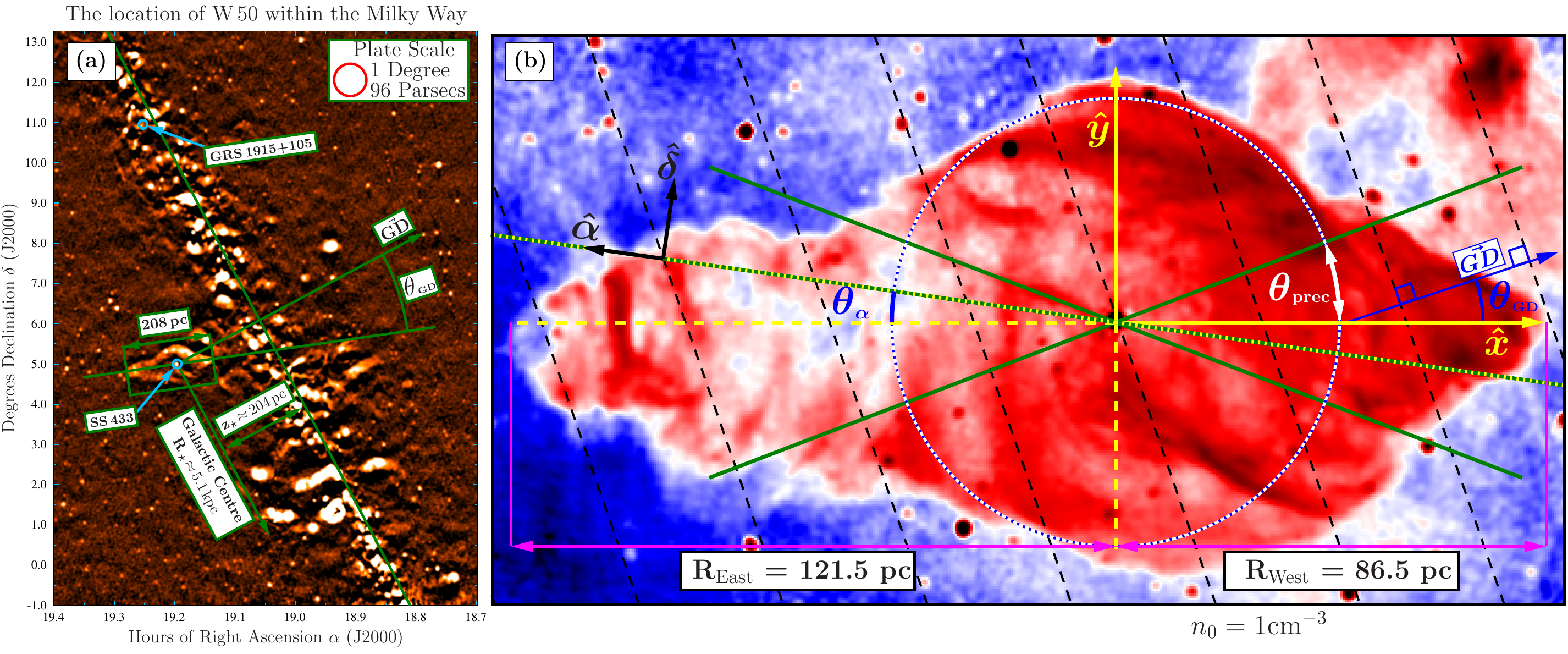}
\caption{This figure describes the geometry of the SS\,433-W\,50 system.  (a) This mosaic was created using archival data from the GBT6 survey \citep{gbt6}, showing the location of W\,50 with respect to the Milky Way Galaxy plane.  The vector $\vec{GD}$ points normally towards the Galactic disc from SS\,433, and ~$\theta_{\small \rm GD}$ = 19.7\dg~ is the angle this vector subtends with the mean jet axis.  The plate scale in (a) applies to objects at the same distance as SS\,433 of 5.5kpc (1 arcmin $\simeq$ 1.6 pc).   (b) The famous VLA mosaic of W\,50 provided by \citet{Dubner} has been transformed here such that the x-axis is coincident with the mean axis of SS\,433's jet precession cone.  The solid green lines indicate the current jet precession cone in SS\,433 which subtends an angle of $\theta_{\rm prec}$ with the mean jet axis.  SS433's western jet-cone points in the direction $+\hat{x}$, and the eastern jet-cone in $-\hat{x}$.  In this frame the Right-Ascension ordinate $\hat{\alpha}$ makes an angle of 172\dg~with the x-axis ~($\theta_{\rm \alpha}$ = 8\dg).  An example of the Galactic density profile used in our model with $\zd$=40\,pc is indicated by the dashed black lines.  Adjacent dashed black lines are separated by ($\zd\ln{2}$) parsec corresponding to changes by a factor of 2 in density, and the density profile is normalised to $n_{\rm 0}$=1 particle cm$^{-3}$ at SS\,433.  Another microquasar, GRS\,1915+105 is also present in the field of view of (a) and its coordinates are indicated.  However, GRS\,1915+105 does not seem to be encapsulated within a well defined large-scale nebulous structure in the way that SS\,433 is with W\,50.}
\flab{allscales}
\end{minipage}
\end{figure*}

\section{Introduction to the SS\,433-W\,50 complex}

The paradigmatic microquasar SS\,433 has become widely known since its discovery as the first stellar source of relativistic jets within our Galaxy \citep{Margon,Fabian,Milgrom}.  After three decades of research\footnote{At the time of writing, of order $10^{3}$ research articles have been devoted to the study of SS\,433 and its environs.}, the system has been widely identified as a high mass X-ray binary (HMXB) system.  The masses of the constituent bodies have been estimated using a variety of techniques \citep{Crampton,Odorico,Collins,Gies,Hillwig,Fuchs,Lopez,Blundell4} yielding some varied results.  In general the observations seem to favour the scenario of a black hole in orbit around a more massive companion star \citep{Fabrika}.  Recently, the presence of a circumbinary ring around SS\,433 has been observed via optical spectroscopy \citep{Blundell4} and also near infra-red spectroscopy \citep{Perez1}, and these observations determine the mass internal to the ring to be 40M$_{\odot}$, of which 16M$_{\odot}$ is attributable to the compact object and its accretion disc.  Thus, it is believed that SS\,433 hosts a black hole (BH) rather than a neutron star.  The BH's accretion disc is presumably fed by gas from a strong stellar wind or Roche-lobe-like overflow from the companion star with a rate of \hbox{$7 \times 10^{-6} \lesssim \dot{M}_{\rm transfer} \lesssim 4 \times 10^{-4} \,{\rm M_{\odot} \,year^{-1}}$} \citep{King}, accompanied by two oppositely directed relativistic jets which are thought to eject material at a rate \hbox{$\dot{M}_{\rm jet} \gtrsim$ 10$^{-6}$ M$_{\odot}$ year$^{-1}$} \citep{Begelman} into the surrounding interstellar medium.  The disc-jet system exhibits precessional motion, which is roughly described by the kinematic model of SS\,433 \citep{Milgrom,Abell} and is readily observed via the periodic Doppler shifting of optical emission lines from the jets.   By fitting the kinematic model to spectroscopic data spanning 20 years, \citet{Eikenberry} determined the following best-fit kinematic parameters:  \hbox{$t_{\rm prec}$ = (162.375 $\pm$ 0.011)} days for the precession period, $\beta_{\rm jet}$ = (0.2647 $\pm$ 0.0008) for the jet speed, \hbox{$\theta_{\rm prec}$ = (20\dg .92 $\pm$ 0\dg.08)} for the precession cone half-angle, and \hbox{$\theta_{\rm inc}$ = (78\dg.05 $\pm$ 0\dg.05)} for the inclination of the mean jet axis to our line of sight, although \citet{blundell2005} found that these properties vary with time.  \citet{Goranskii} conducted a multi-colour photometric study of SS\,433 over a similar time period of 20 years, and used Fourier analysis to extract the orbital period of \hbox{$P_{\rm orb}$ = 13.082} days, and a nutation period of \hbox{$P_{\rm nut}$ = 6.28} days.  

In the radio regime, more than two complete precessional cycles of the jets have been spatially resolved \citep{Blundell2}, providing confirmation of the jet speed as $\sim$0.26c and enabling an independent calculation of the distance to SS\,433 as \hbox{$d_{\rm SS433}$ = 5.5 $\pm$ 0.2 kpc}.  Using Very Long Baseline Interferometry (VLBI), the jet has been resolved at milliarcsec scales into individual fireballs or {\em bolides} of ejecta \citep{Vermeulen,Paragi}, and the presence of an equatorial {\em ``ruff''} outflow from the accretion disc has become apparent \citep{paragi_ruff,Blundell1}.  Furthermore, recent optical spectroscopic studies have revealed that these discrete fireballs of jet ejecta are optically thin, and expanding in their own rest frame at approximately 1\% of the jet speed \citep{Blundell3}.    

SS\,433 is embedded within the W\,50 nebula, as shown in the radio mosaic of \citet{Dubner}.  The remarkable phenomenology of W\,50  (\fref{allscales}b) spans 208\,pc along its major axis, featuring a circular region which is believed to be the remains of an expanding supernova remnant (SNR) shell, with two elongated lobes to the east and west of the circular shell.  The circular shell is centred upon the position of SS\,433 with an offset of just 5 arcminutes \citep{Lockman}.
The distance to W\,50 is difficult to constrain using the kinematics of the HI gas within the nebula because of its extent and inhomogeneity, and due to the high level of turbulent confusion along the line of sight near the Galactic plane.  However, recent radio observations of HI in absorption and emission now confirm that the distance to W\,50 is consistent with 5.5\,kpc \citep{Lockman}, which places it at the locality of SS\,433.  Hence we consider the details of the local environments of SS\,433 and W\,50, which are the same distance from us, such that their mutual evolutions are connected.  

\section{Prevalent questions pertaining to the SS\,433-W\,50 complex}
\slab{questions}

The detail revealed by the famous radio mosaic of W\,50 \citep{Dubner} presented a number of puzzling questions concerning its formation.  We endeavour to address these issues through our models and will later discuss each one (see \sref{snrresults}, \sref{jetresults} and \sref{snrjets}).

\subsection{East-west lobe asymmetry and annular structure}
\slab{asymmetry}

The lobes share an axis of symmetry with SS\,433's mean jet axis, and exhibit a fascinating annular structure (most noticeable in the eastern lobe, \fref{allscales}b).  This has led to the hypothesis that SS\,433's jets are responsible for the formation of W\,50's elongated lobes.  There is a noticeable asymmetry in the extents of the east and west lobes of W\,50 (\fref{allscales}b), and this can be characterised by the ratio
\begin{equation}
R_{\rm ratio} = \frac{R_{\rm East}}{R_{\rm West}} = 1.40.
\end{equation}
Due to the inclination of the jet-axis with respect to our line of sight, a small geometrical correction $\sin \theta_{\rm inc}\approx1$ is implicit in the measured values $R_{\rm East}$ and $R_{\rm West}$ (but not in the ratio $R_{\rm ratio}$), where $\theta_{\rm inc}\approx80^{\circ}$ \citep{hjellming1981,Eikenberry}. As a result the physical extent of each lobe is larger than measured, by a factor 1/$\sin  \theta_{\rm inc}$.  This asymmetry between the eastern and western lobes is thought to be due to the exponential ISM density gradient towards the Milky Way disc, and the validity of this theory is explored in \sref{jetlobes} of this paper.

\subsection{SS\,433's latency period}
\slab{latencyperiod}

We can calculate an {\em in vacuo} jet travel time required for the extent of the eastern lobe to reach $R_{\rm East}$ based on the jet speed:
\begin{equation}
t_{\rm vacuo} = \frac{R_{\rm East}}{\beta_{\rm jet}\, c} \simeq 1500 ~{\rm years}
\end{equation}
where $t_{\rm vacuo}$ is only a lower-limit estimate for the actual jet travel time; in reality the travel-time would be longer due to retardation of the jets in the presence of the ISM.  This estimate is noticeably smaller than the time taken for a typical SNR to reach a radius of 45 parsecs (the radius of the circular region in \fref{allscales}b), suggesting that SS\,433's jets were initiated relatively recently in the history of W\,50.  We refer to the period after the supernova explosion of SS\,433's progenitor and before the ignition of jets, as SS\,433's ``latency'' period.  The reason for this latency period is currently not well understood, and we endeavour to estimate the duration of this period in \sref{conclusions} in order to constrain the possible scenarios.

\subsection{SS\,433's jet persistence and precession-cone angle}
\slab{outbursts}
SS\,433 is often noted as being somewhat unique among the microquasar class due to the apparent persistence of its radio jets.  Other microquasars typically exhibit a more intermittent behaviour, temporarily increasing in luminosity with each episodic jet outburst, before returning to a state of relative quiescence.  These jet ejection episodes in microquasars characteristically last for one or more weeks, and recur every one or more years.  Successive jet outbursts can be observed to happen with substantially different properties, for example in 1997 a flare from Cygnus X-3 was observed with jet speed $v_{\rm 1997}\geq0.81c$ and precession cone angle $\psi_{\rm 1997}\lesssim 12^{\circ}$ \citep{mioduszewski2001}, while a subsequent flare was observed in 2001 with jet speed $v_{\rm 2001}=0.63c$ and precession cone angle $\psi_{\rm 2001}=2.4^{\circ}$ \citep{mj2004}.  We must consider the possibility that SS\,433's jets are not constantly active, but rather that SS\,433 also exhibits jet outburst episodes, albeit of a much longer duration\footnote{In other words, we are currently witnessing (and have been since its discovery) SS\,433's most recent outburst of jet activity.} than has been observed in other known microquasars.

\subsection{Previous simulation work  pertaining to W\,50}

The range in physical sizes involved in the SS\,433-W\,50 system makes it an extremely difficult object to model numerically, however several authors have confronted the challenge of simulating the jet-SNR interaction (notably \citet{Kochanek,velazquez2000,zavala2008}).  Although a more detailed comparison will be given in \sref{grandslam}, we briefly introduce some of the early approaches here.

An early hydrodynamical study of SS\,433's jets was carried out by \citet{Kochanek}, who considered various forms of hollow cylindrical and conical jets, as well as filled conical jet models.  The effect of ``self interaction'' of the jets with previous ejecta and with the jet cocoon itself is mentioned in their paper, and in several of the references therein.  They use an axisymmetric model, and report that their hollow conical jets are neither recollimated, nor refocused.  They discuss how their results are in disagreement with a model for jet recollimation proposed by \citet{eichler1983}, in which precessing jets were suggested to experience a focusing effect caused by the pressure from the ambient ISM through which they propagate.

\citet{velazquez2000} recognised the physical attractiveness of the jet-SNR interaction scenario.  They modelled the W\,50 nebula by instantaneously imposing an analytical Sedov profile upon the ISM to mimic the supernova explosion, upon which the relativistic jets of SS\,433 are later incident.  Due to the computational constraints at the time, they modelled only one quadrant of the SNR in two dimensions, with the simplification of a scaled-down version of W\,50 of radius 10pc and a uniform density background ISM.  The system evolution was closely followed until 2700 years after the jets were switched on, such that the simulated jet-SNR lobe extends to approximately 33pc from the SNR origin.  They follow the example of \citet{clarke1989} who describe a neat model for computing the synchrotron emission from 2-D simulations, by assuming a toroidal magnetic field configuration about the jet axis, and revolving the resulting emission about $2\,\pi$ radians to create the projected 3-D emission.  The hydrodynamics of the jet-SNR was further studied in the 3-d simulations of \citet{zavala2008}, who invoke both hollow conical jets to simulate the precession in SS\,433.  They also conduct a small number of tests to investigate the physical parameters such as jet mass-loss rate and the radius of the SNR at the time when the jets switch on.  The possibility of a time-variable jet precession angle is also tested in their investigations, and their results show a qualitatively similar morphology to that observed in W\,50, in the sense that a spherical SNR shell is present with lateral extensions in the form of jet lobes, of arbitrary size.

Despite the successes of these pioneering works, there are several invalid assumptions made in the models which have significant effects upon the results, and can be improved upon.  The physical significance of the morphology present in W\,50 has (so far) not been adequately addressed with relevance to the pressing questions (see \sref{questions}).  The inconsistencies in these previous works will be discussed more fully in section \sref{grandslam}, and we endeavour to show that it is possible to produce a model of the SS\,433-W\,50 interaction that is consistent with observational data, without imposing any unreasonable assumptions.

We emphasise the importance of including relevant observational data in simulation models, such as the actual sizes of the SNR and jet lobes, if we are to provide useful constraints on the system. 
We incorporate all of these observational constraints (with the exception of orbital motion) into our hydrodynamical model, to give an accurate representation of the dynamics of SS\,433's jets.
Using the observed parameters as a basis for our models of the SS\,433-W\,50 complex, we consider three possible evolutionary scenarios in detail in order to see if we can explain the origins of this extremely large and rare nebula.  The results of these three scenarios are discussed in detail in \sref{snrresults}, \sref{jetresults} and \sref{snrjets}, but their mechanisms can be summarised as follows:  (\rmnum{1}) a supernova blast wave expanding in the presence of a realistic exponential baryon density background appropriate to SS\,433's location in the Milky Way (\rmnum{2}) the interaction of SS\,433's jets with the ISM only (no supernova), and finally (\rmnum{3}) the interaction of SS\,433's jets with an evolved SNR shell from (\rmnum{1}).

However, our intention is not solely to replicate the morphology of SS\,433-W\,50 complex, but to explore the effects of important astrophysical parameters such as the jet mass-loss rate, supernova blast energy, and the ambient ISM density, which are among input parameters to our supernova and jet models.  The distinctiveness of the W\,50 nebula produced through these interactions, allows us to place useful constraints upon these system parameters.

\section{Our Numerical Simulations}

To conduct our simulations, we utilise the well-known hydrodynamic FLASH\footnote{Courtesy of the University of Chicago Center for Astrophysical Thermonuclear Flashes: \phref{blue}{http://flash.uchicago.edu/website/home/}{$http://flash.uchicago.edu/website/home/$}}  code \citep{Fryxell}.  The code implements an adaptive mesh refinement (AMR) method, a higher order Godunov method known as the Piecewise Parabolic Method (PPM) \citep{Colella}.  FLASH is noted for its flexibility, and several case-specific enhancements were made by the authors; most notably the implementation of an exponential Galactic density background and our supernova explosion and jet model modules, which are described in this section.

High resolution VLBI imaging of the central engine in SS\,433 reveals jet structure on scales of a few tens of  AU, a factor of $10^6$ or so smaller than the extent of W\,50\footnote{The site of jet-launch in SS\,433 is thought to be very much closer to the black hole than can be revealed by the resolution of modern telescopes.  Based on the Schwarzschild radius for a $M_{\odot}$ black hole, base of the jets in SS\,433 are likely to be of order $10^{14}$ times closer to the singularity than the lobes of W\,50.}.  The need for a high resolution representation of the core region of SS\,433, and the large-scale field of view (FOV) necessary to enclose W\,50, makes for a challenging computational problem.  

\fref{allscales} shows the location and orientation of W\,50 within the Milky Way and defines the axes of the simulation domain; the $\hat{x}$ and $\hat{y}$ axes lie in the plane of the page, and the $\hat{z}$ axis points out of the page, with SS\,433 at the coordinate origin.  To maximise our resolution in the regions of interest, we simulate a thin slice through the centre of the SS\,433-W\,50 complex, with grid dimensions (L$_{\rm x}$\,$\times$\,L$_{\rm y}$) = (230\,$\times$\,115) parsecs.  The grid is initially split into ($n_{\rm xb}\times n_{\rm yb}$) {\em blocks} before any ``refinement'' of the grid takes place, and it is these {\em blocks} that are subject to refinement.  In order to ensure that all elements on the grid are square rather than rectangular, we set $n_{\rm xb}$=16 and $n_{\rm yb}$=8.  Our chosen domain size ensures that neither jets nor the SNR reach the domain boundaries, such that outflow boundary conditions can be suitably applied.  Our limiting resolution depends on the maximum level of refinement ($\eta_{\rm max}$) allowed, such that: 

\begin{equation}
~~\delta x ~=~ \frac{L_{\rm x}}{[\,n_{\rm xb}\,\times\,2^{\eta_{\rm max}-1}\,]} ~=~ \delta y ~=~ \frac{L_{\rm y}}{[\,n_{\rm yb}\,\times\,2^{\eta_{\rm max}-1}\,]}.
\elab{resolution}
\end{equation}

The domain thickness in the $\hat{z}$ direction is not constant; each cell across the simulation domain has a thickness in $\hat{z}$ equal to its length in $\hat{x}$ (which depends on the cell refinement level), and so the domain is effectively 2D (once cell thick in $\hat{z}$).

We set the maximum refinement level to $\eta_{\rm max}$ = 10, resulting in a maximum resolution of $\delta x$ = $\delta y$ = 0.028\,pc = 8.66$\times 10^{14}$ metres on our grid, or $\simeq$ 1 arcsecond on the sky for objects at a distance of $d_{\rm W\,50} = 5.5$kpc.  This resolution corresponds to about three times the synthesised beam-width from the famous VLA observation of SS\,433's corkscrew jets \citep{Blundell2}.  For simplicity we refer to the maximally refined cells as pixels (cells for which $\eta_{\rm cell} = \eta_{\rm max}$), and due to the nature of AMR, every cell in the simulation domain therefore has sides equal to an integer power of 2 times the pixel size.  Where present, radiative cooling refers to the standard cooling functions provided with FLASH2.5.  The standard radiative cooling in FLASH2.5 assumes an optically thin plasma, where the cooling function $\Lambda(T)$ is based upon models of the energy losses from the transition region of the solar corona \citep{rosner1978} and the chromosphere \citep{peres1982}, and the resulting cooling function is given as a piecewise-power law approximation (refer to the user guide available on the FLASH website).  \corx{We acknowledge that the cooling function adopted by FLASH2.5 assumes that the ions and electrons are in thermodynamic equilibrium, which is not accurate for young supernova remnants.  However, this is not a major issue for our simulations, as we are interested in the large scale, evolved structure of the W\,50 nebula.  A comparison of both cooled and uncooled simulations is given in \fref{snrstats}.}

\subsection{Setting the scene for SS433's progenitor}

We make the simple but effective assumption of approximately cosmological abundances of Hydrogen (90\%) and Helium (10\%) by number\footnote{We note that this corresponds to approximately 70\% Hydrogen and 30\% Helium by mass, which is overabundant in Helium by a few percent with respect to the cosmologically accepted primordial values.}, such that \,$\mu=1.3$\, is the mean mass per particle in atomic mass units.  We use a Galactic exponential density profile for the ISM, adapted from \citet{Binney} of the form:

\begin{equation}
~~\rho_{\rm _{ISM}}(R,z) = \rho_{\rm 0} \,\exp{\Big( -\frac{R_{\rm m}}{R_{\rm d}} -\frac{R}{R_{\rm d}} -\frac{z}{\zd}\Big)}
\elab{rhoprof}
\end{equation}
where the constant $R_{\rm m}$ = 4kpc, ~$R_{\rm d}$ = 5.4\,kpc is the scale length of the stellar disc, and $\zd$ = 40pc is the scale height from the Galaxy disc.  The prefactor $\rho_{\rm 0}$ is determined from a normalisation condition that the density at the location of SS\,433 is:

\begin{equation}
 \rho_{\rm 0} = \rho_{_{\rm \,ISM\,}}(R_{\rm _{SS433}},z_{\rm _{SS433}}) = n_{\rm0}\,\mu \,m_{\rm p}
 \elab{rho0}
\end{equation}
where $m_{\rm p}$ is the proton mass, and $n_{\rm 0}$ is a parameter (for $m_{\rm p}$ in CGS units this defaults to 1 particle per cc, as shown in \fref{allscales}b).  The cylindrical coordinates\footnote{This is calculated using $d_{\odot\rm -GC}$ = 8\,kpc as the Sun-Galaxy centre distance, $d_{\odot \rm-SS433}$ = 5.5\,kpc as the distance from the Sun to SS\,433, which has Galactic coordinates ($l_{\rm SS433}$,\,$b_{\rm SS433}$)  = (39.7$^{\circ}$, $-$2.24$^{\circ}$).}  of SS\,433 relative to the Galactic centre are approximately $R_{\star}=5.1$\,kpc and $z_{\star}=204$\,pc as shown in \fref{allscales}.

Ordinarily, a gravitational term should be invoked to maintain static equilibrium of the background density.  In this scenario, the gravitational centre of the Milky Way galaxy resides far outside of the simulated field of view.  To simplify this we allow the background temperature profile to have an inverse behaviour to that of the background density in order to keep the thermal pressure constant, which artificially creates hydrostatic equilibrium in the background medium.  Preliminary tests were carried out on an unperturbed background to verify the stability of this, and the system remained static over a period of 10$^{6}$ years of simulated time.  Therefore, over the timescales involved in these simulations, we can reliably conclude that subsequent hydrodynamic motion is solely driven by the supernova and/or jet interaction.

The subsequent sections \sref{SNR} and \sref{JETS}, are concerned with the specifics of the supernova and jet models respectively, and the results are explored in \sref{snrresults}, \sref{jetresults} and \sref{snrjets}.

\section{Modelling the Supernova Explosion}
\slab{SNR}

Typically in supernova simulations, a self-similar Sedov-Taylor analytic solution for a strong blast wave is adopted and used to initialise the density, temperature, pressure and velocities of the supernova explosion.  This has the advantage that, for a given background density and explosion energy, the radius of the advancing shockwave behaves predictably with time, according to the Sedov-Taylor \citep{sedov1959, taylor1950} blastwave relation:

\begin{equation}
~~r_{\rm blast}(t) ~~ = ~~ \alpha ~\Big( \frac{E_{\rm blast}}{\rho_{0}} \Big)^{ \frac{1}{5}}\,t^{\frac{2}{5}}.
\elab{sedov}
\end{equation}

However, this description is appropriate only during the Sedov phase of a SNR blast wave, when the mass swept-up by the supernova shock front is comparable to, or greater than the ejected mass from the progenitor.  For a sensible progenitor mass ejection and an ISM particle ($\mu=1.3$) density in the range \hbox{$n_{\rm 0}$\,=\,0.1-1  cm$^{-3}$}, the Sedov phase occurs when the SNR radius falls in the approximate range $10^{0} - 10^{1}$ pc, which is quite large compared to our resolution capability.  Thus, we model the supernova in the pre-Sedov {\em free-expansion} phase, and there are two main advantages to this.

First, it is difficult to guess the initial blast energy of the supernova explosion which may have produced W\,50.  To compensate for this unknown we test two values for the initial kinetic energy of the gas ejected through the supernova blast.  We adopt the notation whereby supernova model $SNR_{\rm50}$ has an initial blast kinetic energy of \hbox{$E_{50}=10^{50}$ ergs} and model $SNR_{\rm 51}$ has \hbox{$E_{51}=10^{51}$ ergs}, and together these encompass a reasonable range of supernova cases.  A purely Sedov-like expansion with thse parameters requires $10^{5}$ years (for the $E_{51}$ case) to $10^{6}$ years (for the $E_{50}$ case) to expand to a radius \hbox{$R_{\rm W\,50}$ = $R_{\rm 45}$ = 45pc} (the current size of W\,50's circular shell region, as per \fref{allscales}b).  The free-expansion phase is much more rapid initially, but eventually the system hydrodynamically establishes a Sedov-like behaviour (explained in \sref{snrresults}).  This method reduces the required time for the SNR radius to reach $R_{\rm 45}$ to something much more reasonable, in agreement with observational predictions of other old SNRs.  The free-expansion SNR reaches a maximum radial expansion speed of [1-2]\%\,{\em c} and as such a relativistic treatment is not critical at this stage.

Second, the scene of our supernova explosion is set within an exponential density background from the Milky Way.  As such it is beneficial to initialise the supernova explosion as early as possible in order to fully investigate the effects of the non-uniform ISM upon the SNR as it evolves and sweeps up ambient material.  This is particularly relevant in order to study the East-West asymmetry present in W\,50.

\subsection{Implementing the Pre-Sedov Supernova Explosion}
\slab{presedov}
We define $R_{\rm Sedov}$ to be the radius at which a sphere of ISM density\footnote{The background density can be considered constant across the small scales relevant here.} $\rho_{\rm 0}$ contains the mass $M_{\rm ej}$ equal to that ejected from the progenitor, and the free-expansion phase is valid for $R(t) < R_{\rm Sedov}$.  The initial radius $R_{\rm i}$ of the supernova blast must satisfy two conditions: (i) $R_{\rm i}$ should be as small as possible in order to capture the very earliest evolution of the explosion within a non-uniform ISM, (ii) $R_{\rm i}$ must be large enough so that the initial supernova blast adequately resembles an unpixellated circle on the simulation grid, to prevent propagation of unwanted artefacts commonly encountered when modelling using cartesian grids.  We parameterise the initial radius as a function of the Sedov radius by
\begin{equation}
 R_{\rm i} = f_{\rm r}\,R_{\rm Sedov} = f_{\rm r}\,\Big(\frac{3M_{\rm ej}}{4\pi \rho_{\rm 0}}\Big)^{\frac{1}{3}}
 \elab{initialr}
\end{equation}
and we find that $f_{\rm r} = 0.25$ satisfies the two conditions above, for a mass ejection of $M_{\rm ej} = 5{\rm M_{\odot}}$ and an ambient density of $n_{\rm 0}=1$ particle ${\rm cm^{-3}}$.  

The free-fall supernova blastwave is initialised as an over-dense disc\footnote{The kinetic energy injected into the SNR disc in the simulation domain is a factor  $\chi_{\rm D}$\,=\,(4\,$R_{\rm i}$\,/\,3\,$\delta x$) of $E_{\rm blast}$, due to the ratio of the volume of a sphere of radius $r_{\rm i}$ with the volume of the disc of thickness $\delta x$ used in the simulation domain.} of radius $R_{\rm i}$ about the explosion epicentre ($x_{\rm c},y_{\rm c}$), with density
\begin{equation}
\rho_{\rm i} = f_{\rm r}^{-3}\,\rho_{\rm 0}.
\elab{initialrho}
\end{equation}
All blocks within the radius $a_{\rm ref} = c_{\rm ref}\,R_{\rm i}$ of ($x_{\rm c},y_{\rm c}$) are maximally refined, so that the grid region in the vicinity of the supernova blast is always described by our maximum resolution.  Some preliminary tests showed $c_{\rm ref}=1.2$ to be sufficient.  
Using the reasonable approximation that the supernova blast energy is initially entirely kinetic, we explore two velocity profiles; the first velocity profile assigns a constant speed $v_{\rm 1}$ to all cells across the supernova initialisation zone, with radial unit vector direction:
\begin{equation}
v_{\rm 1}\hat{r} = \sqrt{\frac{2\,E_{\rm blast}}{M_{\rm ej}}}\,\hat{r}.
\elab{vprof1}
\end{equation}
More realistically, very young SNR can be modelled as uniformly expanding spheres ({\em homologous expansion}), whereby (within the sphere) the radial velocity of the expanding gas is proportional to the radial distance from the sphere centre.  Thus, the second velocity profile allows the gas speed to increase linearly (a technique also used by \citet{jun1996}) from zero at the explosion epicentre, to a maximum value $k_{\rm v}v_{\rm 1}$ at $R_{\rm i}$:
\begin{equation}
v_{\rm 2}(r)\,\hat{r} = k_{\rm v}\,\frac{v_{\rm 1}}{R_{\rm i}}\,\hat{r}
\elab{vprof2}
\end{equation}
where $k_{\rm v}=\sqrt{10/3}$ is the necessary constant to maintain the total kinetic energy as $E_{\rm blast}$.
Although preliminary tests indicate that these profiles are indistinguishable from one another at large scales (see \sref{snrresults}), the second velocity profile minimises the effects of internal shocks and instabilities to which the supernova explosion is susceptible at very early times.  As such, the latter velocity profile provides a neat method to ensure that the ensuing SNR is more closely symmetrical.  This is particularly important when considering that jets will later be incident upon structures internal to the SNR, and will help to minimise stochastic effects which could affect the jet propagation in an asymmetric manner, voiding any meaningful comparison between the east and west jet evolution.

We follow the supernova evolution until the blastwave radius reaches approximately 44\,pc, allowing for some further expansion of the SNR towards $R_{\rm 45}$ once the jets are switched on.

\begin{figure*}
\centering
\begin{minipage}{\textwidth}
\centering
\mygraphics{width=\textwidth}{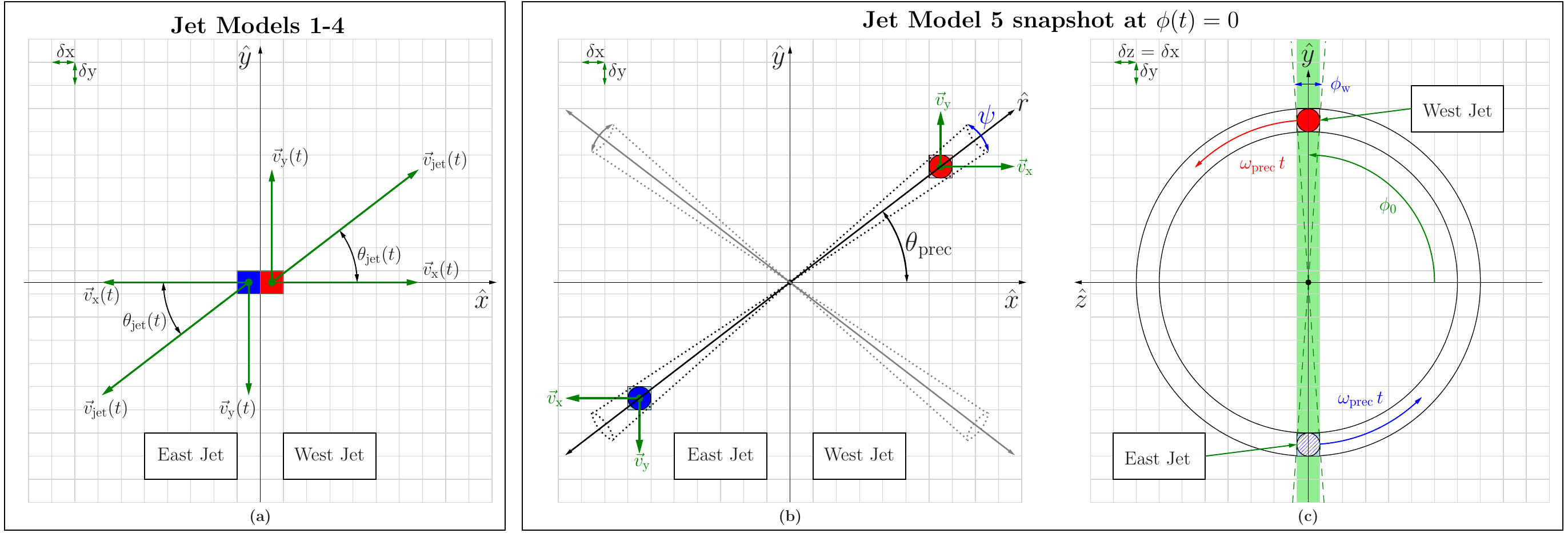}
\caption{A graphical summary of the jet models.  We use blue and red colours to denote SS\,433's east (mostly blueshifted) and west (mostly redshifted) jets respectively, from the point of view of an observer at Earth.  This use of colour is purely to differentiate between the east and west jets in the figure, and we remind the reader that our simulated jets spend an equal amount of time in approach and recession in the simulation frame, due to the axis of symmetry.  The green section in (c) indicates the simulation slice to which the precessing jets contribute plasma.  In the geometry of (a) and (b), the $\hat{z}$ axis points out of the page, and makes an angle ($90-\theta_{\rm inc}$) with the line-of-sight towards Earth.}
\flab{jetmods}
\end{minipage}
\end{figure*}

\section{Modelling SS\,433's Jets}
\slab{JETS}

We consider the behaviour of SS\,433's jet at the current epoch, which consists of a jet precessing around a cone.  We stress however, that this need not have always been the case\footnote{In fact, it is much more likely that SS433's jets behaved somewhat differently in the past, see Section \sref{jetresults}.}, and we must allow for the possible evolution with time of SS433's jet precession since the formation of the compact object.  We discuss various interesting jet geometries in turn, in the subsections that follow.  The parameters which govern jet behaviour in our models are summarised in Table \tref{jetparms}, along with their default values in convenient units (note that SI units are assumed wherever these parameters appear in equation form).

\begin{table} 
\begin{minipage}{\columnwidth}
\centering
\caption{Summary of the jet parameters.}
\begin{tabular}{c c c}
\hline\hline
{\bf Symbol} & {\bf Description} & {\bf Constraint } \\
\hline
$\dot{M}_{\rm jet}$ & Jet mass loss rate  & $\gtrsim$ 10$^{-6}$ M$_{\odot}$ yr$^{-1}$ \\
${v}_{\rm jet}$ & Average jet speed  & 0.2647\,{\bf c} \\
$\theta_{\rm prec}$ & Jet cone half-angle &20.92\dg\\
$t_{\rm prec}$ & Jet precession period  & 162.375 days \\
$\omega_{\rm prec}$ & Jet angular speed  &  3.8695$\times$10$^{-2}$ rads s$^{-1}$ \\
$T_{\rm jet}$ & Jet temperature  & 10$^{4}$\, Kelvin  \\
\hline\hline
\end{tabular} 
\tlab{jetparms}
\end{minipage}
\end{table}
\medskip

Note that $\dot{M}_{\rm jet}$ and $T_{\rm jet}$ are poorly constrained.  Since the mass loss in the jets is not well known, we investigate three orders of magnitude in $\dot{M}_{\rm jet}$ ranging from a relatively weak jet $10^{\rm -6}$ M$_{\odot}~{\rm yr.}^{\rm -1}$, an intermediate jet of $10^{\rm -5}$ M$_{\odot}~{\rm yr}^{\rm -1}$, and a strong jet of $10^{\rm -4}$ M$_{\odot}~{\rm yr.}^{\rm -1}$.  The jet temperature $T_{\rm jet}$ is also not very well known, but the prominent H$_{\rm \alpha}$ emission in the jets suggest the temperature is at least $10^{\rm 4}$ Kelvin.  An estimate of SS\,433's jet temperature is given in \citet{Fabrika} as $2\times 10^{4}$\,K.

Numerically, we introduce SS\,433's jets by including relevant source terms in the Euler hydrodynamic conservation equations.  The conservation equations governing SS\,433's jets can be written as:

\begin{equation}
\pd{\,\rho}{t} + \Pdiv{\rho\pvec{v}} ~=~ \dot{\rho}_{jet}
\elab{euler1}
\end{equation}

\begin{equation}
\pd{\,\rho \pvec{v}}{t} + \Pdiv{\rho\pvec{v} \otimes \pvec{v}} ~=~ -\Pgrad{P} + \dot{\rho}_{\rm jet}\,{\pvec{v}}_{\rm jet}
\elab{euler2}
\end{equation}

\begin{equation}
\pd{\,\rho \epsilon}{t} + \Pdiv{\rho \epsilon\,\pvec{v}} ~=~ -\Pdiv{P\,\pvec{v}} - \dot{\epsilon}_{\rm rad} + \dot{\epsilon}_{\rm jet}
\elab{euler3}
\end{equation}
where $\rho$, v, p, and $\epsilon$ are the mass density, velocity, thermal pressure and total energy density respectively, and $\dot{\epsilon}_{\rm rad}$ is the radiated power (per unit volume) due to radiative cooling.  A non-relativistic treatment for the jet gas is adopted, i.e. the ISM and jet plasmas are assumed to have the same adiabatic index $\gamma$ = 5/3.  The density, temperature, and velocities of the jet injection region are reset to (\,$\rho_{\rm jet}\,,\,T_{\rm jet}\,,\,v_{\rm x}\,,\,v_{\rm y}$\,) for each time-step ($v_{\rm z}$ = 0 everywhere), and the density $\rho_{\rm jet}$ is a model-dependant function of $\dot{M}_{\rm jet}$.  

The structure created by a jet in its surrounding environment is primarily determined by the jet velocity components $v_{\rm x}$, $v_{\rm y}$ and $v_{\rm z}$ which are also model dependant, but their most generalised forms follow the kinematic model\footnote{Note that here the kinematic model is adapted slightly because SS\,433's mean-jet-axis is the x-axis of the simulation domain.  This explains the absence of the inclination angle dependence $\cos\theta_{\rm inc}$ and $\sin\theta_{\rm inc}$ in this representation.} of SS\,433 such that:

\begin{equation}
\vec{v}_{x}\big(\theta_{\rm jet}(t),\phi(t)\big) = v_{\rm jet}\,\cos\big(\theta_{\rm jet}(t) + C_{\theta}\big) ~\vec{\hat{x}}
\elab{velx}
\end{equation}
\begin{equation}
\vec{v}_{y}\big(\theta_{\rm jet}(t),\phi(t)\big) = v_{\rm jet}\,\sin\big(\theta_{\rm jet}(t) + C_{\theta}\big)\cos\phi(t) ~\vec{\hat{y}}
\elab{vely}
\end{equation}
\begin{equation}
\vec{v}_{z}\big(\theta_{\rm jet}(t),\phi(t)\big) = v_{\rm jet}\,\sin\big(\theta_{\rm jet}(t) + C_{\theta}\big)\sin\phi(t) ~\vec{\hat{z}}
\elab{velz}
\end{equation}
\begin{equation}
\phi(t) = \omega_{\rm prec}\,(t - t_{\rm 0}) + \phi_{\rm 0}
\elab{phioft}
\end{equation}
\begin{equation}
C_{\rm \theta} =
\Bigg\{
\begin{array}{l} 
0 ~~~~~for~the~Western~jet \\
 \\ 
\pi ~~~~~for~the~Eastern~jet
\end{array} 
\elab{ctheta}
\end{equation}
where $\phi(t)$ is the precession phase angle at time t, and $t_{\rm 0}$ is the time when the jets switched on.  Throughout this paper we use the definition whereby precessional phase zero occurs when the red-most (western) jet is maximally\footnote{In the simulations, the maximum redshifts are equal for the east and west jets, due to the axis of symmetry used.} red-shifted (i.e. the blue-most jet is maximally blue-shifted).  SS\,433 precesses in a clockwise fashion\footnote{The right hand grip rule is useful here: gripping the x-axis with the right hand, with the thumb in the +$\hat{x}$ direction, the curled fingers indicate the precession direction for the west jet.} when looking along the direction of the western jet from the viewpoint of SS\,433.  Consequently, if the phase angle starts from zero at $t=t_{\rm 0}$ then the offset $\phi_{\rm 0}=-\pi/2$ ensures that $\phi(t)=n\pi/2$ when the jets are maximally Doppler shifted, and $\phi(t)=n\pi$ when the jets display only the transverse Doppler shift (in the xy-plane), where $\{n\,\in\,\mathbb Z^{*}\}$ is the revolution number.

To allow for the possibility that the precessional motion in SS\,433 has changed during its lifetime as an X-ray binary, either smoothly or as a discontinuous process (c.f. episodic jet outbursts in \cygx\ mentioned in \sref{outbursts}), we introduce into our models some simple time-dependence in the precession cone angle:
\begin{equation}
\theta_{\rm jet}(t) = \theta_{\rm 0} + (t-t_{\rm 0})\,\dot{\theta}
\elab{jettheta}
\end{equation}
which can be made constant by setting $\dot{\theta}$=0.

The jet density depends on the jet mass loss rate $\dot{M}_{\rm jet}$ and some model-dependant geometry, and the internal pressure is calculated from the density:
\begin{equation}
\rho_{\rm jet} = \alpha_{\rm geom} \,\dot{M}_{\rm jet}
\elab{jetrho}
\end{equation}
\begin{equation}
P_{\rm jet} = \frac{\rho_{\rm jet}\,k_{\rm B}\,T_{\rm jet}}{\mu\,m_{p}}
\elab{presjet}
\end{equation}
where $\alpha_{\rm geom}$ has units of $time/volume$.  The total power per unit volume for the jets has kinetic and thermal components:
\begin{equation}
\dot{\epsilon}_{\rm jet} = \frac{\dot{P}_{\rm jet}}{(\gamma - 1)} + \frac{1}{2}\,\dot{\rho}_{\rm jet}\,v^{2}_{\rm jet} ~=~ \dot{\rho}_{\rm jet} \,\big(\, \frac{k_{\rm B}\,T_{\rm jet}}{\mu\,m_p \,(\gamma -1)} + \frac{v^{2}_{\rm jet}}{2} \,\big).
\elab{edensjet}
\end{equation}

We explore 5 different jet initialisation models, and later discuss the relative successes of each.  Models 1 to 4 are based upon the jet motion described in equations \eref{velx} through \eref{jettheta} and depicted in \fref{jetmods}a, whereas model 5 attempts to mimic SS\,433's discrete ballistic jet ejecta, as depicted in \fref{jetmods}b.

\subsection{{\bf Jet Model 1} - Static Cylindrical Jet}
Our most basic jet model consists of a static (non-precessing) cylindrical jet along the x-axis.  This is a special case of the general equations \eref{velx} through \eref{jettheta}, created by setting $\theta_{\rm 0}$ = 0, $\dot{\theta}$ = 0, and $\phi(t)$ = 0, such that the velocity components are reduced to $\vec{v_{\rm x}}=v_{\rm jet}\cos{C_{\rm \theta}}\,\hat{x}$ and $\vec{v_{\rm y}}=\vec{v_{\rm z}} = 0$.  

The density of the jet is simply a product of three factors: the mass loss rate $\dot{M}_{\rm jet}$ in the jets, the (resolution-dependant) pixel crossing time $t_{\rm \times}=\delta x/v_{\rm jet}$, and the model-dependant jet volume \hbox{($n_{\rm jx}\,n_{\rm jy}\,n_{\rm jz}  \,\delta x^{3}$)}, where ($n_{\rm jx}$,\,$n_{\rm jy}$,\,$n_{\rm jz}$) are the number of pixels constituting the jet initialisation region in the respective ($\hat{x}$,\,$\hat{y}$,\,$\hat{z}$) directions.  This issue can be simplified by choosing the initial jet size (as in \fref{jetmods}a) to be 1 cubic pixel ($n_{\rm jx}$\,=\,$n_{\rm jy}$\,=\,$n_{\rm jz}$\,=\,1).  This makes the most physical sense, since the pixel size is very much larger than the jet structure revealed by VLBI imaging, which is in turn very much larger than a few Schwarzschild radii of the black hole in SS\,433.  Thus, the density in this model is given according to equation \eref{jetrho} with:
\begin{equation}
\alpha_{\rm geom} = \frac{1}{2\,v_{\rm jet}\, \delta x^{2}}
\elab{geom1}
\end{equation}
where the above summed over both jets will give the required mass loss rate per unit length (one pixel) in $\hat{x}$.  

Simulations based upon this jet model were performed for each of the three $M_{\rm jet}$ values, both with and without the presence of an evolved SNR.  The simulation runs for this model are also used to quantify the east-west jet asymmetry in W\,50 by constraining the density normalisation $n_{\rm 0}$ and Galaxy scale-height $\zd$ in equation \eref{rhoprof} appropriate at the location of SS\,433 in our Galaxy, and these results are discussed in \sref{jetlobes}.

\subsection{{\bf Jet Model 2} - Conical Phase-shiftng Jet:\, $(\,\dot{\theta}_{\rm jet} = 0\,)$}
This model invokes a time-averaged representation of SS\,433's precessing conical jet, termed \emph{phase-shifting}.  As with model 1, the east and west jets are each just 1 cubic pixel in size.  This model uses present-day observed parameters describing SS\,433's precession at the current epoch, and assumes the jets have not changed throughout their history.

The computational time-step is adjusted so that precession phase-space is sampled $n_{\rm s}$ times per precession period:
\begin{equation}
\Delta t_{\rm step} = \frac{t_{\rm prec}}{n_{\rm s}}
\elab{dtmod2}
\end{equation}
\begin{equation}
\Delta \phi_{\rm step} = \frac{2\pi}{n_{\rm s}}
\elab{dphimod2}
\end{equation}
If we define a phase-tolerance as $\phi_{\rm tol}$=$\Delta \phi_{\rm step}/2$, then the precessing jet contributes\footnote{The jet pixels are set according to the jet values ($\rho_{\rm jet},T_{\rm jet},\vec{v}_{\rm jet}$) rather than the background ISM values.} to the simulation domain when the phase $\phi(t)$ falls within $\pm \phi_{\rm tol}$ of 0 (for the upper-half of the plane) or $\pi$ (for the lower-half).  When this happens, the phase $\phi(t)$ is rounded to the nearest of either of (0,\,$\pi$), and this is equivalent to introducing an error of $\pm$($50/n_{\rm s}$) percent to the precession period (hence the term \emph{phase-shifting}).

The velocities are initialised by setting \hbox{$\theta_{\rm 0}$ = $\theta_{\rm prec}$}, \hbox{$\dot{\theta}$ = 0} and \hbox{$\phi(t)$ = $\omega_{\rm prec}t$} in equations \eref{velx} through \eref{jettheta}, where \hbox{$\omega_{\rm prec}$ = 2$\pi$/$t_{\rm prec}$}.  Assuming that the amount of mass outflowing in the jets in one precession period can be written as $\dot{M}_{\rm jet}\,t_{\rm prec}$, the density is set according to equation \eref{jetrho} with:

\begin{equation}
\alpha_{\rm geom} = \frac{t_{\rm prec}}{2\,n_{\rm s}\,\delta x^{3}}.
\elab{geom2}
\end{equation}
We choose $n_{\rm s}=20$ as a compromise value to prevent impractical inflation of the computational time, whilst limiting the error in precessional period to just $\pm2.5\%$.

\subsection{{\bf Jet Model 3} - Conical Phase-shiftng Jet:\, $(\,\dot{\theta}_{\rm jet} > 0\,)$}
This model is similar to Jet Model 2 in that it follows the same precession \emph{phase-shifting} model, but with the exception that the precession cone angle is allowed to increase linearly with time.  The basis for this model comes from the discrepancy between the opening angle of the jet precession cone $\theta_{\rm prec}$ currently observed in SS\,433, and the angle subtended by the lobes of W\,50 at the central source SS\,433.  The green lines in \fref{allscales}b represent the current jet precession angle projected out to the extent of W\,50.  Had SS\,433's jets penetrated the lobes of W\,50 at this angle, we would expect quite a different morphology to result.  
To investigate the effects of a smoothly evolving jet using equation \eref{jettheta}, the initial half-cone angle is set to $\theta_{\rm 0}=0$ and the derivative becomes
\begin{equation}
\dot{\theta} = \frac{\theta_{\rm prec} - \theta_{\rm 0}}{t_{\rm jets}}
\elab{thetadot}
\end{equation}
such that the cone half-angle increases linearly to the currently observed value, where $t_{\rm jets}$ is the time taken for the jets to reach the extent of the lobes in W\,50.  The jet density is as described in model 2.

\subsection{{\bf Jet Model 4} - Conical Phase-shiftng Episodic Jet }
In this model we investigate the possibility that SS\,433's jet activity is discontinuous.  This is inspired by the microquasar \cygx, which exhibits episodic jet outbursts, each with quite different jet speeds and precession cone angles.  To emulate this behaviour, we use the same \emph{phase-shifting} model and jet density as described in jet model 2, but the outbursts are created by using $\dot{\theta}=0$ in equation \eref{jettheta} and setting $\theta_{\rm 0}$ according to:
\begin{equation}
\theta_{\rm 0} =
\Bigg\{
\begin{array}{l} 
\theta_{\rm 1}  ~~~~~for~~~ t_{\rm 1} < t < t_{\rm 2}\\ 
\theta_{\rm 2}  ~~~~~for~~~ t_{\rm 3} < t < t_{\rm 4}\\
\theta_{\rm 3}  ~~~~~for~~~ t_{\rm 5} < t < t_{\rm 6}.
\end{array} 
\elab{mod4thetas}
\end{equation}

\subsection{{\bf Jet Model 5} - Fireball Jet Model}

The fireball model was inspired by the spectacular animation\footnote{See: \phref{blue}{http://www.nrao.edu/pr/2004/ss433/ss433.movie.gif}{$http://www.nrao.edu/pr/2004/ss433/ss433.movie.gif$}} showing the discrete fireball-like nature of SS\,433's jets from the work of \citet{mioduszewski2003}.  The optical spectroscopy observations of \citet{Blundell3} estimate the expansion speed $u_{\rm e}$ of the jet fireballs in the innermost regions of SS\,433 as approximately $u_{\rm e} \simeq$ 1\% of the jet speed.  Using this estimate of the fireball expansion rate, and by assuming that the initial fireball size upon ejection is negligible compared to our resolution, we approximate the initial jet collimation angle as \hbox{$\psi$ $\simeq$ $u_{\rm e}/v_{\rm jet}$ $\simeq$ 0.01} radians.  Thus, it is possible to calculate the distance from the origin at which the fireballs will have expanded to a size comparable to 1 simulation pixel, such that we can reasonably resolve the discrete ejecta seen in VLBI observations on the simulation grid.  For approximately spherical fireballs of radius $R_{\rm fb}$ then \hbox{(2R$_{\rm fb}$ = r$_{\rm fb}\,\psi$ = $\delta x$)} gives the appropriate fireball displacement \,$r_{\rm fb}$\, along its trajectory $\hat{r}$.  Hence we can write the coordinates of the fireball pixel in each  quadrant as:
\begin{equation}
x_{\rm fb} ~=~ x_{\rm 0}~\pm~\frac{\delta x \,\cos \theta_{\rm prec}}{\psi} ~=~ x_{\rm 0}~\pm~93.4\,\delta x
\elab{fbx}
\end{equation}
  \begin{equation}
y_{\rm fb} ~=~ y_{\rm 0}~\pm~\frac{\delta x \,\sin \theta_{\rm prec}}{\psi} ~=~ y_{\rm 0}~\pm~35.7\, \delta x
\elab{fby}
\end{equation}
and there are 4 such fireball pixels; one in each quadrant of the $xy$-plane as indicated in \fref{jetmods}b. 

In three dimensions the jet sweeps out part of a corkscrew structure in space over one precession period, however our domain is effectively a slice one pixel ($\delta x$) thick in the $\hat{z}$ direction.  It is useful to define the quantity $f_{\rm fb}$ which is the angle subtended by one fireball pixel about the x-axis, as a fraction of one whole precessional period (2$\pi$ radians):
\begin{equation}
f_{\rm fb} = \frac{\delta x}{2\pi r_{\rm fb}} = \frac{\psi}{2\pi \sin \theta_{\rm prec}}
\elab{ffb}
\end{equation}
where $\delta x$ is the pixel size and effective width of the simulation slice in $\hat{z}$.  We now introduce a user-defined constant $\kappa_{\rm fb}$ (equal to 1 by default), which allows the user to establish a compromise between computation time and resolution of the fireballs.  The precessing jet only contributes plasma when it sweeps through the simulation domain slice.  In phase-space this spans half of the phase-width \hbox{$\phi_{\rm w}$ = 2$\pi\,\kappa_{\rm fb}\,f_{\rm fb}$} each side of the $xy$-plane, or \hbox{($0\pm\pi \kappa_{\rm fb}f_{\rm fb}$)} for the fireballs in the northern (upper-half) plane, and \hbox{($\pi\pm \pi \kappa_{\rm fb}f_{\rm fb}$)} for the southern (lower-half) plane.  This phase range is specified by the criterion:
\begin{equation}
\cos^{2}\phi(t) > \cos^{2}\phi_{\rm w}
\elab{phicriteria}
\end{equation}

The computational time-step $\Delta t_{\rm step}$ must be reduced for this model, to ensure that the jet phase $\phi (t)$ evaluated at $t_{\rm n}$ and \hbox{$t_{\rm n} + \Delta t_{\rm step}$} does not cause the jet to skip the $xy$-plane altogether.  The precession crossing time for the jet to sweep through the simulation slice is equal to  \hbox{$\Delta t_{\rm cross}$ = $\kappa_{\rm fb}\,f_{\rm fb}\,t_{\rm prec}$}.  To ensure adequate sampling of the precession period, the computational time-step in this model needs to be equal to or smaller than the precession crossing time:
\begin{equation}
\Delta t_{\rm step} \lesssim \Delta t_{\rm cross} = \frac{ \kappa_{\rm fb}\,\psi\,t_{\rm prec}}{2\pi \sin\theta_{\rm prec}} \approx 0.72 ~{\rm days}
\elab{fireballtstep}
\end{equation}
which (with $\kappa_{\rm fb}=1$) is typically a few orders of magnitude smaller than other dynamical timescales on the simulation grid.  The jet plasma density is given according to equation \eref{jetrho} with:
\begin{equation}
\alpha_{\rm geom} = \frac{\kappa_{\rm fb} \,f_{\rm fb}\,t_{\rm prec}}{2\,\delta x^{3}}.
\elab{geom5}
\end{equation}
The jet plasma is initialised for the fireball pixels according to:
\begin{equation}
\big(\rho\,,T\,,\vec{v}\big) =
\Bigg\{
\begin{array}{l} 
\big(\rho_{\rm jet}\,,T_{\rm jet}\,,\vec{v}_{\rm jet}\,\big)   ~~~~~{\rm for}~~~\cos^{2}\phi(t) > \cos^{2}\phi_{\rm w}\\ 
 \\ 
\big(\rho_{\rm ISM}\,,T_{\rm ISM}\,,0\big) ~~~ {\rm otherwise}.
\end{array} 
\elab{fballinit}
\end{equation}

\begin{table*} 
\begin{minipage}{17cm}
\centering
\caption{Summary of our preliminary simulations testing the SNR parameters.  The first six runs test the background profiles and radiative cooling.  The last four simulation runs are variations of SNR6, where meaningful astrophysical parameters are tested.}
\begin{tabular}{c c c}
\hline\hline 
{\bf Shortname} & {\bf Description} & {\bf Parameter Details} \\
\hline
SNR1 & Supernova only & SNR defaults and:~$\rho$-profile\,=\,0, v-prof\,=\,0, Cooling=off \\
SNR2 & Supernova only & SNR defaults and:~$\rho$-profile\,=\,1, v-prof\,=\,0, Cooling=off \\
SNR3 & Supernova only & SNR defaults and:~$\rho$-profile\,=\,0, v-prof\,=\,1, Cooling=off \\
SNR4 & Supernova only & SNR defaults and:~$\rho$-profile\,=\,1, v-prof\,=\,1, Cooling=off \\
SNR5 & Supernova only & SNR defaults and:~$\rho$-profile\,=\,0, v-prof\,=\,1, Cooling=on \\
SNR6 & Supernova only & SNR defaults and:~$\rho$-profile\,=\,1, v-prof\,=\,1, Cooling=on \\
SNR6b & Supernova only & As SNR6 but:~ $E_{\rm blast}=10^{\rm 50}\,$ergs \\
SNR6c & Supernova only & As SNR6 but:~ $n_{\rm 0}=0.1\,{\rm cm}^{\rm -3}$ \\
SNR6d & Supernova only & As SNR6 but:~ $M_{\rm ej}=10$\,M$_{\odot}$ \\
SNR6e & Supernova only & As SNR6 but:~ $T_{\rm bg}=10^{\rm 2}\,$ Kelvin\\
\hline
\multicolumn{3}{c}{SNR default parameters: $M_{\rm ej}=5$\,M$_{\odot}$,\, $n_{\rm 0}=1\,{\rm cm}^{\rm -3}$,\, $E_{\rm blast}=10^{\rm 51}\,$ergs,\, $T_{\rm bg}=10^{\rm 4}$\,Kelvin,\, $\zd=40$\,pc} \\
\hline
\hline 
\end{tabular} 
\tlab{snrsimref}
\end{minipage}
\end{table*}
\medskip
If the criterion of equation \eref{phicriteria} is met (thus signifying a jet intersection with the $xy$-plane), the precessional phase must be very close to either of (0,\,$\pi$) and we round $\phi(t)$ to the nearest of these.  This ensures that tiny residual $v_{\rm z}$ values are set to zero when the fireball velocities are initialised:
\begin{equation}
\left( 
\begin{array}{c} 
v_{\rm x} \\ 
v_{\rm y}\big(\phi(t)\big) \\ 
v_{\rm z}\big(\phi(t)\big)
\end{array} \right) = 
\left( 
\begin{array}{c} 
\cos(\theta_{\rm prec}+C_{\rm \theta}) \\ 
\sin(\theta_{\rm prec}+C_{\rm \theta})\cos \phi(t) \\ 
\sin(\theta_{\rm prec}+C_{\rm \theta})\sin \phi(t)
\end{array} \right) ~v_{\rm jet}
\elab{fballvelos}
\end{equation}
where $C_{\rm \theta}$ has the usual meaning of zero for the west jet ($x>0$) and $\pi$ for the east jet ($x<0$).

\subsection{Computation and Data Analysis}
A total of 32 preliminary simulations were conducted for this paper, all of which were performed using the Oxford Astrophysics {\em Glamdring} cluster.  The characteristics of each simulation are outlined in Tables \tref{snrsimref} and \tref{jetsims}, and for convenience we will be refer to each simulation by its shortname given therein.

All of the scripts used to analyse the data from our simulations were coded using the Perl Data Language\footnote{The Perl Data Language (PDL) has been developed by K. Glazebrook, J. Brinchmann, J. Cerney, C. DeForest, D. Hunt, T. Jenness, T. Luka, R. Schwebel, and C. Soeller and can be obtained from http://pdl.perl.org.}.  The simulations employ the Hierarchical Data Format (HDF5) option for data output in FLASH, which is appropriate for use with AMR.  Using code written in PDL, each HDF5 file is converted to a uniform grid for easier analysis, and stored as (archived) Flexible Image Transport System (FITS) files with appropriate header information relevant to each simulation.

\section{SNR: Results \& Discussion}
\slab{snrresults}

A series of 10 preliminary supernova tests were performed to investigate the effects of the key physical parameters governing the evolution of the SNR (see Table \tref{snrsimref}).  The supernova blast energy $E_{\rm blast}$, the background density $n_{\rm 0}$ at SS\,433, and the mass ejected by the progenitor $M_{\rm ej}$ were among the parameters tested, as well as the effects of radiative cooling and changing the background density profile from uniform to exponential.  The velocity profiles described in section \sref{presedov} are also tested, but they do not have a meaningful interpretation.  

A sample of snapshot images from two of these SNR simulations, \SNR{6c} (left column) and \SNR{6} (right column) are shown at various time intervals in \fref{snrfig}.  The figure illustrates the difference in the SNR evolution when the background density at SS\,433 is varied from $n_0=0.1{\rm cm}^{-3}$ (left column) to $n_0=1{\rm cm}^{-3}$ (right column).  The difference in the initial radii of the supernova blasts from the first image in each column is a consequence of the first density difference by a factor of 10 in equation \eref{initialr}.  The most obvious difference is in the time taken for the SNR to expand to $R_{\rm 45}$, whereby increasing the density by an order of magnitude causes the duration of expansion to increase by more than a factor of 2.  As the supernova shockwave expands into the ISM, the effects of the exponential density gradient become apparent in a number of ways.  First, the density of the west side of the SNR (nearest the Galaxy plane) is noticeably higher than the density at the east side in the evolved shell.  The ratio of the densities $R_{\rho}=\rho_{\rm W}/\rho_{\rm E}$ on opposite sides of the SNR shell reaches a maximum of $R_{\rho}\simeq10$ along the direction normal to the Galactic plane (vector $\vec{GD}$ in \fref{allscales}) in the snapshots at later times, in \fref{snrfig}.  As expected, the east-west density ratio has approximately the same value for both \SNR{6} and \SNR{6c}.  Second, the position of the centre ($x_{\rm 0},y{\rm _0}$) of the SNR quite noticeably shifts as a function of time, with respect to the initial blast epicentre ($x_{\rm c},y{\rm _c}$), moving away from the Galaxy plane, downstream in the density gradient.

\begin{figure*}
\vspace{-0.3cm}
\centering
\begin{minipage}{0.95\textwidth}
\centering
\mygraphics{width=\textwidth}{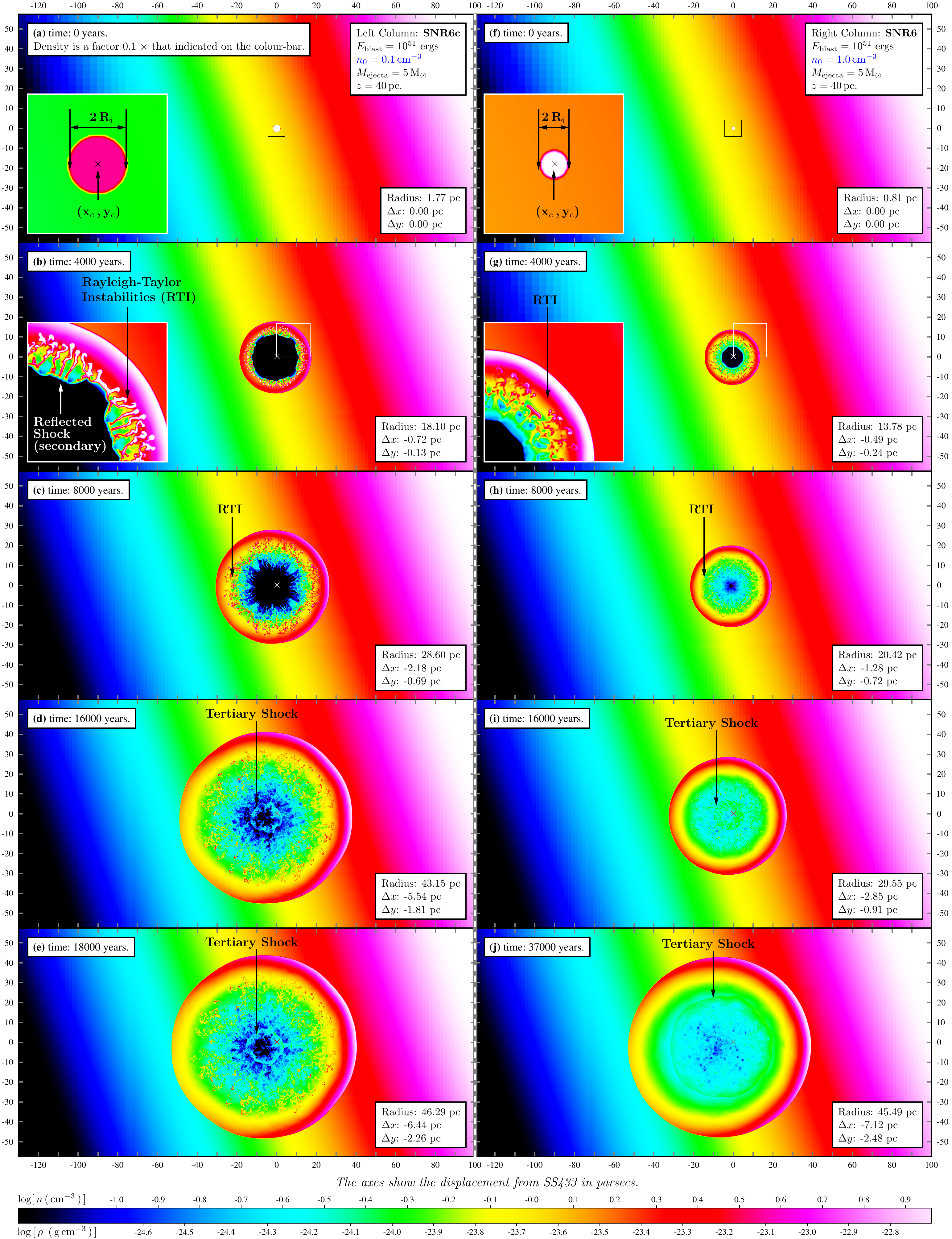}
\caption{The evolution of 10$^{51}$ erg supernova explosions in the presence of the Galaxy density gradient.  Supernova simulations \SNR{6} and \SNR{6c} are shown at 4 time intervals during the SNR evolution, plus the final snapshot when the SNR reaches $\sim$45pc in radius.  The $\times$ symbol indicates the explosion epicentre on each snapshot, and the displacements $\Delta x$ and $\Delta y$ of the SNR from the epicentre ($x_{c},y_{c}$) are also given.  RTI indicates the occurrence of Rayleigh-Taylor instabilities, where lower-density gas from inner radii are flowing into regions of higher density gas at larger radii within the SNR.  The density for the left-hand column (\SNR{6c}) is 10 times lower than the colour bar indicates.  The animations associated with these simulations are available at: \phref{blue}{http://www-astro.physics.ox.ac.uk/~ptg/RESEARCH/research.html}{\small $http://www$-$astro.physics.ox.ac.uk/$$\sim$$ptg/RESEARCH/research.html$}}
\flab{snrfig}
\end{minipage}
\end{figure*}

\subsection{Monitoring the SNR shock front}
\slab{snrshockfront}
To quantify the differences in morphology of the resultant SNR produced by each simulation in Table \tref{snrsimref}, a shock-front detection algorithm  was applied to each dataset.  The coordinates ($x_{\rm i},y_{\rm i}$) of the points lying on the shock-front were fit to a general ellipse of the parametric form:
\begin{equation}
x_{\rm i}(\theta_{\rm i}) = x_{\rm 0} + a\cos(\phi)\cos(\theta_{\rm i}) - b\sin(\phi)\sin(\theta_{\rm i})
\elab{ellipsex}
\end{equation}
\begin{equation}
y_{\rm i}(\theta_{\rm i}) = y_{\rm 0} + a\sin(\phi)\cos(\theta_{\rm i}) + b\cos(\phi)\sin(\theta_{\rm i})
\elab{ellipsey}
\end{equation}
where ($x_{\rm 0},y_{\rm 0}$) is the centre, $\theta_{\rm i}$ is the angle each point ($x_{\rm i},y_{\rm i}$) subtends between the $\hat{x}$ axis and the centre in the anti-clockwise sense, $\phi$ is the angle between the $\hat{x}$ axis and the semi-major axis, and $a$ and $b$ the semi-major and semi-minor axes respectively.  The eccentricity as well as the displacements of the SNR shell centre ($x_{\rm 0},y{\rm _0}$) from the explosion epicentre ($x_{\rm c},y{\rm _c}$) are then easily calculated as:
\begin{equation}
e = \sqrt{1-(b/a)^{2}}
\elab{ellipticity}
\end{equation}
\begin{equation}
\Delta x = x_{\rm c} - x_{\rm 0}
\elab{xshift}
\end{equation}
\begin{equation}
\Delta y = y_{\rm c} - y_{\rm 0}.
\elab{yshift}
\end{equation}

\begin{figure*}
\centering
\vspace{-0.4cm}
\begin{minipage}{0.9\textwidth}
\centering
\mygraphics{width=\textwidth}{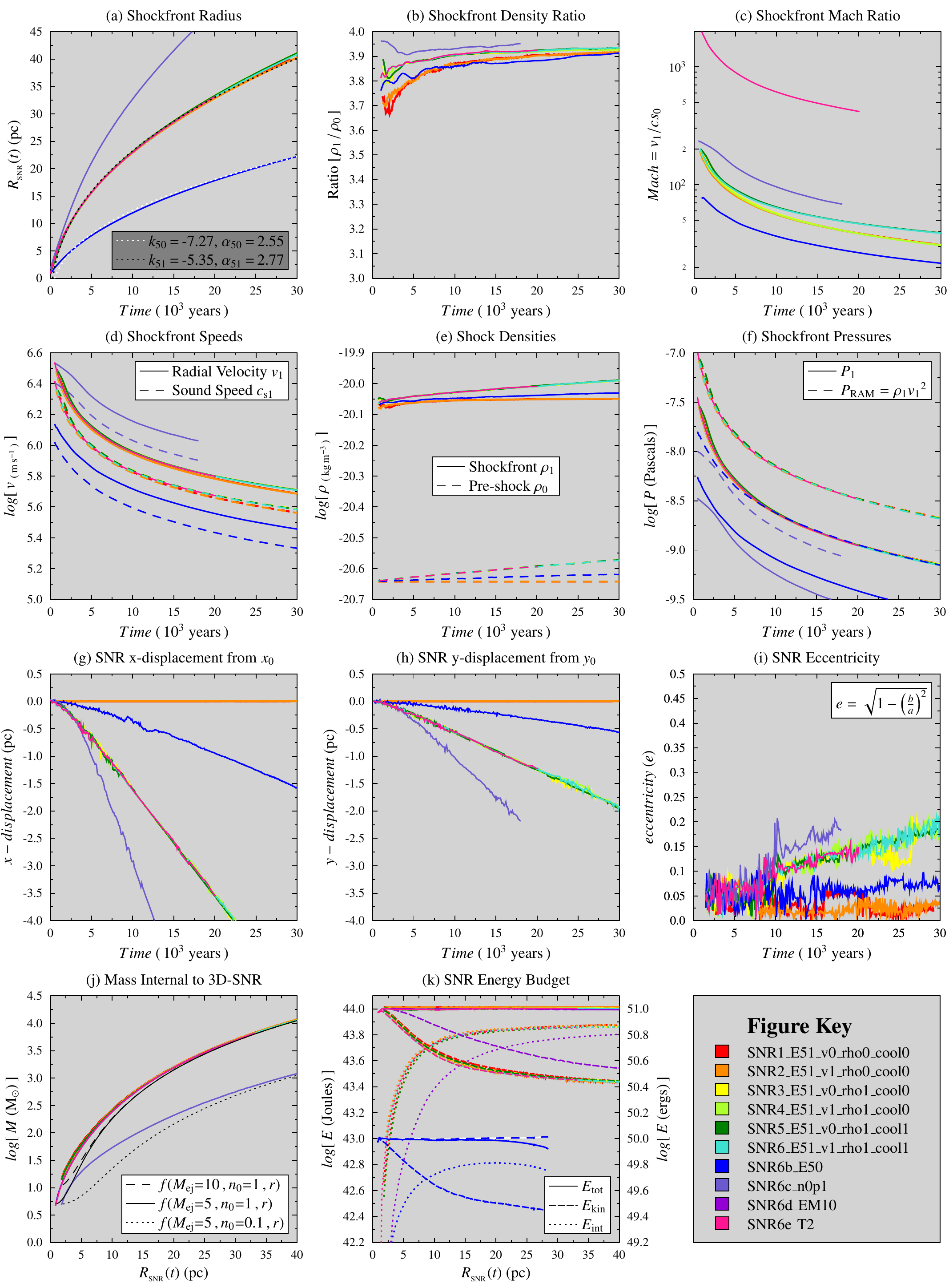}
\caption{A shock-front detection algorithm is used to monitor the size and shape of the SNR, and to sample the density, pressure and velocities of the SNR shell as a function of time.  The figure key summarises the SNR simulation details given Table \tref{snrsimref}, where ``E'' precedes the log of the blast energy in ergs, ``v'' precedes the the velocity profile used (see \sref{presedov}), ``rho'' precedes the background density profile used (0=uniform, 1=exponential), and ``cool'' indicates whether cooling is on (=1) or off (=0).  For simulations \SNR{6b}$-$\SNR{6e}, parameters are only specified where they differ from those used in simulation \SNR{6}. Note that \SNR{6c} in a lower density environment reaches 45\,pc much faster than the other SNRs, and is truncated when the $x$-axis is time.   \SNR{6b} has an order of magnitude lower blast energy than the others, and its shock front dissociates and loses identity before it reaches 30\,pc, such that the shock front cannot be well detected and these data are also truncated.}
\flab{snrstats}
\end{minipage}
\end{figure*}

The physical variables (density, pressure, velocities etc) are sampled at the detected shock-front, and various quantities are plotted as a function of time or SNR radius in \fref{snrstats}.  The shift in centre position of the SNR are plotted in \fref{snrstats}(g) and (h), and it is clear from \fref{snrstats}(i) that although the SNR becomes slightly elliptical in the presence of an exponential density background, the effect is a minor one ($e\lesssim0.2$) compared to the shift of the SNR centres.  In \fref{snrstats}(a) the average radius $R=\sqrt{a^2+b^2}$ of the SNR is plotted as a function of its age, and the dotted lines show the best fit to a $t^{2/5}$ Sedov-like function of the form $k+\alpha\left(\frac{E}{\rho}\right)^{1/5}t^{2/5}$ where $k$ is some offset in parsec.  The shock generated through the supernova explosion is strong as evident from the compression\footnote{Note that the shock-front densities plotted in \fref{snrstats} are actually averaged values from an annulus 1 pixel thick about the detected location of the shock-front, and that densities on the east and west side of the SNR differ by an order of magnitude when the background follows the exponential density profile.} ratio of $\rho_{1}/\rho_{0}\lesssim 4$ across the shock (Fig. \ref{fig:snrstats}b) as well as the ram pressure being higher than the thermal pressure (Fig. \ref{fig:snrstats}f), and the shock speed remains hypersonic (Mach$\gg$5) for the duration of the simulations as shown in Fig. \ref{fig:snrstats}c and \ref{fig:snrstats}d.  In \fref{snrstats}j, the mass contained within an SNR sphere is roughly estimated by averaging the density of the SNR disc of radius $R$ in the simulation domain, and scaling up by the volume of a sphere of the same radius.  An estimate of the mass expected to be swept up by a spherical SNR in a uniform background of density $n_{0}$ is also plotted on \ref{fig:snrstats}j for comparison, with the functional form
\begin{equation}
f(M_{\rm ej},n_{\rm 0},r) = M_{\rm ej} + \frac{4\pi\,\mu\, m_{\rm p}\,n_{\rm 0}}{3}(r^3 - R_{\rm i}^3), ~~~{\rm for} ~r\ge R_{\rm i} 
\elab{massmodel}
\end{equation}
such that the initial mass contained within the young SNR of radius $R_{\rm i}$ is equal to the mass ejected $M_{\rm ej}$ by the progenitor.  This is a crude model, as it makes no attempt to account for the exponential background density or the ISM material contained within $R_{\rm i}$, but it proves to be consistent with the scaled-up mass from the SNR to within a factor of a few.  The final plot (\fref{snrstats}k) tracks the total energy of the system, which is again scaled up by a factor $\chi_{\rm D} = 4\,R_{\rm i}/(3\,\delta x)$, representing the ratio of the volume of a sphere of radius $R_{\rm i}$ to the volume of the initial disc of the same radius, used to initialise the supernova explosion in the simulation domain (see \sref{presedov} for details about this).  The projected energy in a sphere of radius $R_{\rm SNR}(t)$ as shown in \fref{snrstats}k is then:
\begin{equation}
E_{tot} = \chi_{\rm D} \sum_{i}^{n} \left( \frac{P_{\rm i}}{(\gamma - 1)} + \frac{\rho_{\rm i} v_{\rm i}^2}{2} \right)\,dV_{\rm i}
\elab{energybudget}
\end{equation}
representing the discrete summation of the internal and kinetic energies per unit volume, for concentric annuli of volume elements $dV_{\rm i}$ from the centre of the SNR to $R_{\rm SNR}(t)$.  The energy radiated via cooling is tracked as a function of time for each dataset, and radiative loss of energy from the system via cooling is negligible\footnote{Note that the microphysics of emission-line cooling is not yet fully implemented in this version of the code.} for all but one of the SNR simulations. \SNR{6b}, with an order of magnitude lower blast energy than the others, was found to slow down much faster than the other SNRs.  This is expected for a remnant passing through the Sedov phase into the radiative phase in which cooling does become important, and the addition of cumulative radiative losses is included in the total energy budget of the system, as indicated by the dashed blue-white line in \fref{snrstats}k.  \SNR{6b} eventually began to disperse and lose its circular SNR identity, at which point the shock-front detection algorithm failed to detect a well-defined shock and for this reason the blue lines in \fref{snrstats} are truncated at 28\,pc.  The other 9 SNR simulations can be thought of as adiabatic to a good approximation, evidenced by the constancy of the total energy $E_{\rm tot}$ shown by the solid lines in \fref{snrstats}k.  As one might expect, the shock deceleration is less noticeable in the lower density background of \SNR{6c}, and the transfer of kinetic energy into thermal energy is more gradual, as \fref{snrstats}k shows.

\subsection{The displacement of SS\,433 from the SNR centre}
\slab{snrshift}
The shockfront detection algorithm applied to each of the SNR simulations allowed us the fit an ellipse to the shockfront and track the movement of its centre away from the blast epicentre, as a function of time (\fref{snrstats}g and h).  

The movement of a SNR's centre in the presence of a density gradient was mentioned by \citet{chevalier1974}.  The cause of the shifting is not buoyancy, but rather the relative ease with which the eastern side of the SNR can propagate in the lower density ISM, compared to the much higher inertial resistance met by the western side of the SNR in the higher density towards the Galactic plane.  In other words, the shifting is an {\em apparent} effect due to the independent propagation of the east and west shock fronts into different media, and the SNR shell doesn't physically move as a rigid body.

According to \SNR{6} of \fref{snrfig}j, by the time the SNR expands to 40-45\,pc in radius we can expect the centre position to have shifted by $(\Delta S_{\rm x},\Delta S_{\rm y})\simeq(-7.1,-2.5)$ pc or \hbox{$(-4.4,-1.6)$} arcmin.  Note that these shifts are taken from \fref{snrfig}j and therefore are only appropriate to the SNR evolution in an environment where the galactic scale height is $\zd=40$\,pc, and the density normalisation is $n_{\rm 0}=1.0$ cm$^{-3}$.  The displacement values for the case of \SNR{6c} differ only slightly from those of \SNR{6}, but the displacement is likely to be larger in the case of a steeper background density gradient (e.g. $\zd=30$\,pc).  Regardless of the exact value of $\zd$, the apparent shift in SNR-centre will always be in roughly the same direction with respect to the density gradient, i.e. {\em away} from the Galactic disc with -ve $\Delta S_{\rm x}$ and $\Delta S_{\rm y}$.  

We apply a rotation matrix through the clockwise angle $\theta_{\alpha}$ from \fref{allscales}b to determine the equivalent SNR shifts along the ordinates of right ascension and declination:
\begin{equation}
\left( 
\begin{array}{c} 
\Delta S_\alpha \\ \Delta S_{\delta} 
\end{array} 
\right)
=
\left( 
\begin{array}{cc} 
\cos\theta_{\alpha} & \sin\theta_{\alpha}\\
-\sin\theta_{\alpha} & \cos\theta_{\alpha}
\end{array} 
\right)
\left( 
\begin{array}{c} 
-\Delta S_{\rm x} \\ \Delta S_{\rm y}
\end{array} 
\right)
=
\left( 
\begin{array}{c} 
+4.6 \\ -1.0
\end{array} 
\right)
\ {\rm arcmin}
\end{equation}
where the $-\Delta S_{\rm x}$ occurs because the simulation $x$-axis and the right ascension coordinate are almost antiparallel.

Without understanding that the SNR shell hydrodynamically shifts position in the density gradient of the Galaxy, one might mistakenly calculate that SS\,433 is moving away from the SNR centre.  The movement of the SNR must be considered when calculating the age of the black hole by a simple means of off-centre displacement from the SNR centre, divided by SS\,433's apparent proper motion.  We refer the reader to the detailed study of W\,50 described by \citet{Lockman}\footnote{The displacement values are not used here because the article quotes the linear or angular displacement as 4\,pc or 5 arcmin.  However, these length and angle values are discrepant by a factor of 2 for objects at a distance of 5.5\,kpc.} , where the off-centre displacement of determined from X-ray observations \citep{watson1983} is used in conjunction measurements of SS\,433's proper motion (A.\,Mioduszewski \& M.\,Rupen, private communication), to estimate the age of W\,50 of order $10^{5}$ years. 

The correct way to estimate the age $t_{\rm W50}$ of W\,50 by this method must include both the proper motion data from SS\,433 and a contribution from the apparent SNR displacement, such as:
\begin{equation}
\Delta S_{\alpha} = (\,\mu_{\rm \alpha} - v_{\rm \alpha}\,)\,t_{\rm W50}\\
\elab{ageestimate1}
\end{equation}
\begin{equation}
\Delta S_{\delta} = (\,\mu_{\rm \delta} - v_{\rm \delta}\,)\,t_{\rm W50}
\elab{ageestimate2}
\end{equation}
where $\mu_{\alpha}$ and $\mu_{\delta}$ describe SS\,433's proper motion, $v_{\alpha}$ and $v_{\delta}$ are the time-averaged SNR displacement speeds from the simulations according to \hbox{$v_{\alpha}=\Delta S_{\alpha}/t_{\rm SNR}$} and \hbox{$v_{\delta}=\Delta S_{\delta}/t_{\rm SNR}$}, and $t_{\rm SNR}$ is the appropriate simulation time over which the displacement occured.  Equations \eref{ageestimate1} and \eref{ageestimate2} should give the same answer when solved for $t_{\rm W50}$, but this method requires an accurate measurement of $\Delta S_{\alpha}$ and $\Delta S_{\delta}$, which have not yet been obtained from observations. 

\begin{table*}
\begin{minipage}{17cm}
\centering
\caption{Summary of our preliminary simulations testing the jet parameters.  Simulation Jet1 is used as the control group to which Jet2-Jet8 are compared.  Jet9 investigates the effects of an episodic (rather than continuous) jet, and Jet10 demonstrates the fireball model described in the text.  The Galaxy scale height parameter $\zd$ is tested in another series of simulations, as detailed in \fref{jetlobes}.}
\begin{tabular}{c c c l}
\hline\hline 
{\bf Shortname} & {\bf Description} & \multicolumn{2}{c}{{\bf Parameter Details}} \\
\hline
Jet1 & Jets only (Model 1) &  \multicolumn{2}{c}{Jet defaults and:~ $\theta_{\rm 0}=0^{\circ}$,\, $\dot{\theta}=0^{\circ}$\,yr$^{\rm -1}$} \\
Jet2 & Jets only (Model 1) &  \multicolumn{2}{c}{As Jet1 but:~  $\dot{M}_{\rm jet}=10^{-5}\,{\rm M}_{\odot}\,{\rm yr^{\rm -1}}$}\\
Jet3 & Jets only (Model 1) &  \multicolumn{2}{c}{As Jet1 but:~  $\dot{M}_{\rm jet}=10^{-6}\,{\rm M}_{\odot}\,{\rm yr^{\rm -1}}$}\\
Jet4 & Jets only (Model 1) &  \multicolumn{2}{c}{As Jet1 but:~  $n_{\rm 0}=0.1\,{\rm cm}^{\rm -3}$} \\
Jet5 & Jets only (Model 2) &  \multicolumn{2}{c}{As Jet1 but:~  $\theta_{\rm 0}=10^{\circ}$} \\
Jet6 & Jets only (Model 2) &  \multicolumn{2}{c}{As Jet1 but:~  $\theta_{\rm 0}=20^{\circ}$} \\
Jet7 & Jets only (Model 2) &  \multicolumn{2}{c}{As Jet1 but:~  $\theta_{\rm 0}=40^{\circ}$} \\
Jet8 & Jets only (Model 3) &  \multicolumn{2}{c}{As Jet1 but:~  $\dot{\theta}=20^{\circ}/1000\,yr$} \\
\hline
\multirow{3}{*}{Jet9} & \multirow{3}{*}{Jets only (Model 4)} & \multirow{3}{*}{Jet defaults but~ $n_{\rm 0}=0.1\,{\rm cm}^{\rm -3}$ and:}
        & $\theta_{\rm 1}=40^{\circ}$ for $0 < t < 1000$\, yrs   \\
& & & $\theta_{\rm 2}=0^{\circ}$ ~for $1000 < t < 2600$\, yrs \\
& & & $\theta_{\rm 3}=\theta_{\rm prec}$ for $2600 < t < 2700$\, yrs  \\
\hline
Jet10 & Jets only (Model 5) &   \multicolumn{2}{c}{Jet defaults but~ $n_{\rm 0}=0.1\,{\rm cm}^{\rm -3}$~ and~  $\kappa_{\rm fb}=10$}\\
\hline
\multicolumn{4}{c}{Jet default parameters: $\dot{M}_{\rm jet}=10^{-4}\,{\rm M}_{\odot}\,{\rm yr^{\rm -1}}$,\, $n_{\rm 0}=1\,{\rm cm}^{\rm -3}$, $\zd=40$\,pc} \\
\hline
\hline 
\end{tabular} 
\tlab{jetsims}
\end{minipage}
\end{table*}
\medskip

\section{Jets: Results \& Discussion}
\slab{jetresults}

A series of 10 simulations were performed to investigate the effects of the jet kinematics, such as the jet precession cone angle and jet power (influenced through $\dot{M_{\rm jet}}$), and the details of the simulations are given in Table \tref{jetsims}.  Each jet simulation was allowed to continue running until it became obvious that a particular jet model was unlikely to produce the required morphology observed in W\,50.  An image showing the density map for each simulation in its final stage is shown in \fref{jetfig}, and a contour map of the \citet{Dubner} observation of W\,50 has been overlaid upon each density map for comparison. The magnitude of the jet velocity is kept constant at $v_{\rm jet}=0.2647$c for each jet simulation.

\fref{jetfig}a shows \J{1} from Table \tref{jetsims} consists of a simple static cylindrical jet along the $\hat{x}$ axis, with a mass loss rate of $\dot{M}_{\rm jet}=10^{\rm -4}\,{\rm M}_{\odot}\,{\rm yr}^{\rm -1}$ in an exponential Galactic density gradient of scale-height $\zd=40$\,pc and normalised to $n_{0}=1$\,cm$^{-3}$ at SS\,433's location on the grid.  Although the east-west extent of the lobes from \J{1} approximately matches that of W\,50, the eastern jet lobe has reached a displacement of $\sim$125\,pc from SS\,433 which is in excess of W\,50's eastern lobe extent, whilst the simulated western jet lobe is short of W\,50's western lobe, as clearly indicated by the contour line.  This could indicate that the galaxy scale-height used was too small, and this is investigated in \sref{jetlobes}.  As the jets plough through the ISM, they carve out a cavity of lower density which enables the easier propagation of the proceeding jet material.  Referring to the colour scale used in \fref{jetfig}a, a contrast of an order of magnitude or more in density is achieved in some places, between the lower density ISM directly surrounding the eastern jet (blue) and the denser gas of the cavity walls (red) behind the jet shock front.  We can define an average {\rm lobe speed} as the ratio of the average extent of the lobes to the time taken for the jet shock front to reach such an extent:
\begin{equation}
v_{\rm lobes} = \frac{d_{\rm lobes}}{2\, \Delta t}
\elab{lobespeed}
\end{equation}
where $v_{\rm lobes} = 0.148$c for \J{1}, based on 200\,pc as the east-west lobe extent and 2200\,yrs of simulation time.  Note however, that since the jet is ``on'' for the duration for this simulation, that the kinetic energy of the jets is continually replenished, and the preceding jet activity has already evacuated a cavity in the ISM.  This results in the speed of the jet itself remaining very close to $v_{\rm jet}$ until very near to the bow shock at the ends of the jet lobes, and this is described further in \sref{jetkin}.  Upon reaching the ends of the lobes, the jet ejecta decelerates rapidly as momentum and kinetic energy are dissipated into the denser ambient ISM, to drive the shock forward.  As the jet shock-front propagates, it imparts momentum into the transverse ($\hat{y}$) direction, such that the wake of the jet has a well-defined width at a given distance from the jet launch point.  The result is that even a cylindrical jet (zero cone angle) has a substantial width as evident in  \fref{jetfig}a, where clearly the western jet lobe is as wide as (and in some parts, too wide for) W\,50's western lobe.  The width of the jet lobe as a function of distance from SS\,433 is a useful diagnostic in constraining the nature of the jet which sculpted W\,50's lobes, as shown in \fref{lobewidth}.

\J{2} of \fref{jetfig}b is identical to \J{1} but for the mass-loss rate which is an order of magnitude lower at \hbox{$\dot{M}_{\rm jet}=10^{\rm -5}\,{\rm M}_{\odot}\,{\rm yr}^{\rm -1}$}.  The level of jet collimation displayed by \J{1} is not true for \J{2}, and although the jet mass loss rate is still relatively high it is clear that, if allowed to expand to double the size shown in \fref{jetfig}b, the peanut-like lobes created by \J{2} would be much too wide to produce the lobes of W\,50.  In this case the jets have significantly decelerated before moving far from the jet launch point, and the jet loses its well defined shape, bending and buckling and thereby transferring relatively more momentum in the transverse direction than does \J{1}.  The averaged lobe speed for \J{2} is $v_{\rm lobes}=0.058$c. As one might expect, the deceleration and lack of collimation are both even more obvious in \J{3} of \fref{jetfig}c, where the jet mass-loss rate has been reduced by a further order of magnitude to $\dot{M}_{\rm jet}=10^{\rm -6}\,{\rm M}_{\odot}\,{\rm yr}^{\rm -1}$, resulting in an elliptical bubble with average speed $v_{\rm lobes}=0.017$c.

\begin{figure*}
\centering
\vspace{-0.4cm}
\begin{minipage}{0.99\textwidth}
\centering
\mygraphics{width=\textwidth}{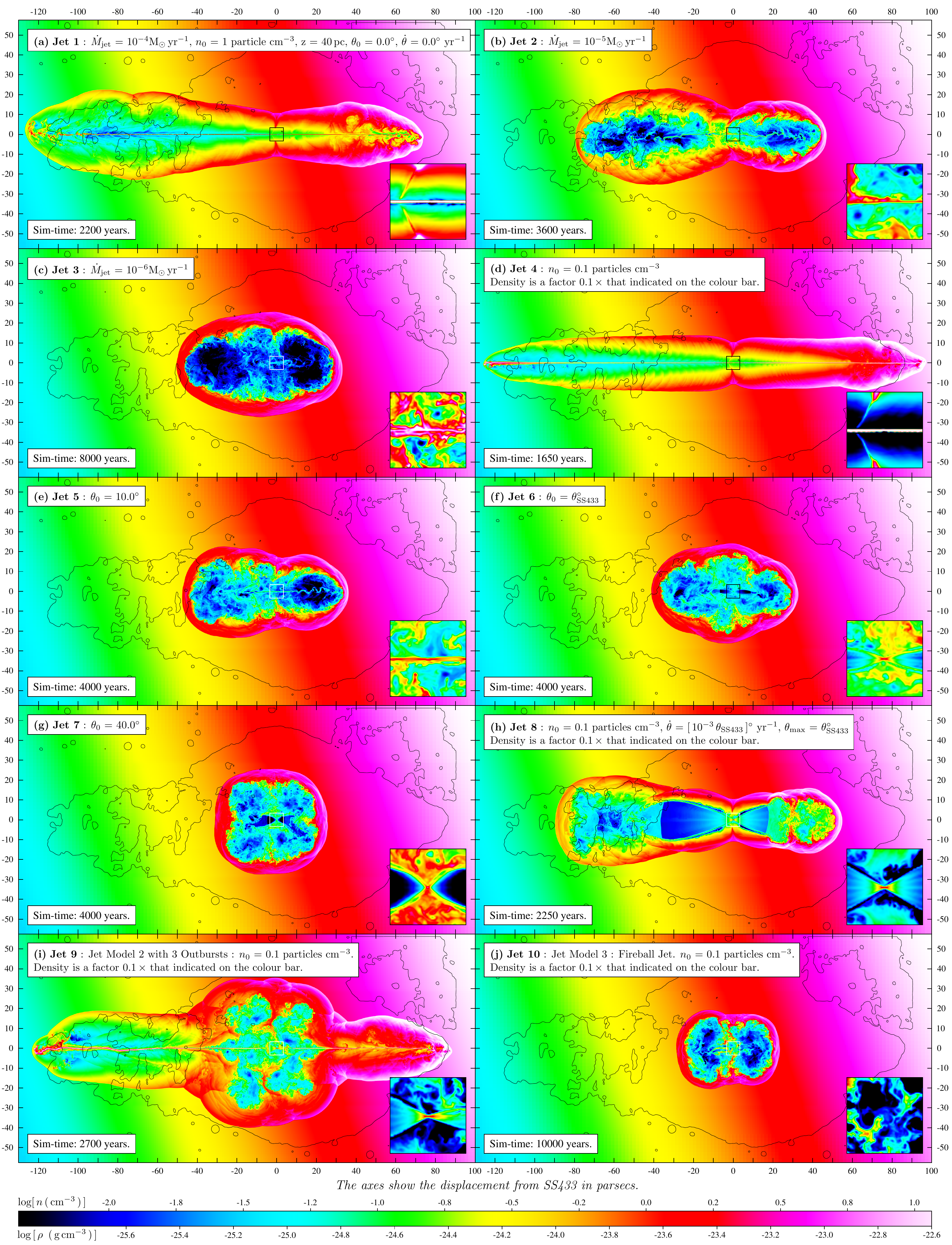}
\caption{Final snapshots are shown of a series of jet simulations demonstrating the effects of relevant jet parameters upon the morphology of the jet cocoon.  Refer to Table \tref{jetsims} for a summary of the differences between these simulations, and \sref{jetresults} for a detailed commentary on the interpretation of each one.  The animations associated with these simulations are available at: \phref{blue}{http://www-astro.physics.ox.ac.uk/~ptg/RESEARCH/research.html}{\small $http://www$-$astro.physics.ox.ac.uk/$$\sim$$ptg/RESEARCH/research.html$}}
\flab{jetfig}
\end{minipage}
\end{figure*}

\J{4} of \fref{jetfig}d has the same jet mass loss rate as \J{1}, but in this case the ambient density normalisation is lower by an order of magnitude: $n_{\rm 0}=0.1$\,cm$^{-3}$.  The relative ease with which this jet propagates through the ISM is apparent both through the level of collimation of the jets and also the jet lobe extents exceed those of W\,50 on each side.  As one might expect, the jets require only a relatively short time (less than 1650 years) to propagate to the full extent of W\,50's lobes ($v_{\rm lobes}=0.219$c), as the background density is lowered and we tend towards the {\em in vacuo} limit described in \sref{latencyperiod}.  The transverse momentum component is reduced relative to the axial component, and jet travel time is short enough that the wake of the simulated jet lobes are thinner than the lobes of W\,50.  This leads to an interesting relationship between the thickness of the jet lobes and the ratio $\dot{M}_{\rm jet}/n_{\rm 0}$.  The effect upon the jet lobes morphology of {\em reducing} $n_{\rm 0}$ by a factor of 10 is equivalent to {\em increasing} $\dot{M}_{\rm jet}$ by the same factor.  For example, repeating \J{4} with $\dot{M}_{\rm jet}$ reduced by a factor of 10, reproduces the same jet morphology as \J{1}, and running \J{1} with an order of magnitude decrease in the background density, reproduces the same jet morphology of \J{4}.  This relationship is described further in \sref{jetlobes}.
  
\J{5} of \fref{jetfig}e utilises Jet Model 2, and has similar parameters to \J{1} with the addition of time-averaged precession of the jet with half-cone angle 10$^{\circ}$, approximately half the jet cone angle $\theta_{\rm prec}$ of SS\,433 current precession state.   Even with this relatively small half-cone angle it is easy to see that the evolved lobes would be much wider than W\,50's lobes, but despite the peanut shaped morphology this jet still produces two distinguishable lobes.  Interestingly \J{6} of \fref{jetfig}f features a rounded central region between two stubby lobes, when the jet half cone angle is set to $\theta_{\rm prec}$.  When the precession half-cone angle is increased to 40$^{\circ}$ is with \J{7} of \fref{jetfig}g, the lobe structure is lost and the jet cavity more closely resembles a circular blown bubble.  Each of the simulations with non-zero precession cone angles display evidence of hydrodynamic refocusing of the jets towards the mean jet axis (for more details see \sref{refocus}).   

To investigate the possibility that SS\,433's precession developed gradually over time, \J{8} of \fref{jetfig}h features a linearly growing jet with half-cone angle starting from zero, and reaching $\theta_{\rm prec}$ after 1000 years.  It was thought that this jet in early stages (with a small precession cone angle) could produce the desired elongated lobes of W\,50, and then at later times when the precession cone angle is larger, the jet would assist in inflating the circular SNR-like region.  Interestingly, the result is quite different because the jet interacts with itself.  As before, the jet ploughs through the ISM and carves out a cavity of low density, providing much lower resistance for the jet that follows at later times.  However, the dense shock-front of ISM gas swept-up by the early jet, also acts to confine the later jet.   Rather than penetrating the cavity walls, the jets interact with the shell at a grazing angle, and are ``guided'' back towards a path of lower resistance.  More details of this effect follow in \sref{refocus}.

\begin{figure}
\centering
\begin{minipage}{0.99\columnwidth}
\centering 
\mygraphics{width=\columnwidth}{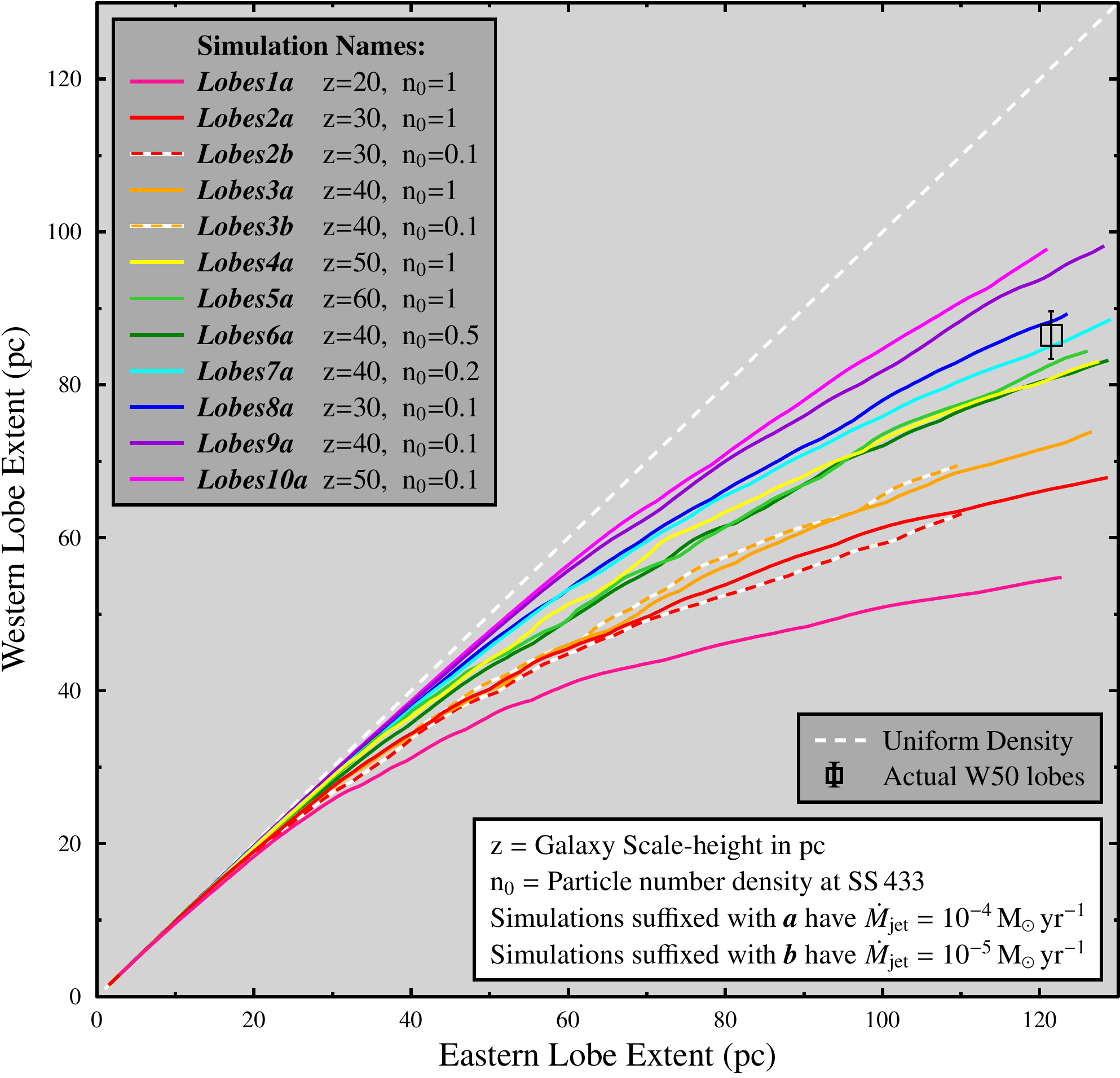}
\caption{We monitor the east and west lobe lengths as a function of time, whilst varying the Galaxy scale-height $\zd$ and ambient density $n_{\rm 0}$.  The distinctive morphology of W\,50 can thus be used to constrain the physical parameters most appropriate to the SS\,433-W\,50 system.  The jet shock-front detection technique can be seen in action at: \hbox{\phref{blue}{http://www-astro.physics.ox.ac.uk/~ptg/RESEARCH/research.html}{\small $http://www$-$astro.physics.ox.ac.uk/$$\sim$$ptg/RESEARCH/research.html$}}}
\flab{jetlobes}
\end{minipage}
\end{figure}

\subsection{Jet lobe propagation: constraining SS\,433's environment}
\slab{jetlobes}
The east-west asymmetry in W\,50's jet lobes, if due to the Galaxy density gradient, provides us with an opportunity to do two things: (i) to investigate hydrodynamically the large-scale effects of jet propagation in a non-uniform background density environment for comparison with observations, and (ii) to constrain the parameters of the Galactic density profile appropriate to the location of SS\,433. 

The purpose of this section is to constrain the parameters $n_{\rm 0}$ and $\zd$ by comparing the jet lobe extents from the simulations to those observed in W\,50.  These two parameters have the following effects:
\begin{enumerate}
\item {\bf Density normalisation $n_{\rm 0}$} - this defines the relative ease with which the jets propagate through the ISM.  Increasing $n_{\rm 0}$ causes greater deceleration of the jets, because both the east and west jets experience greater resistance from the ISM through which they propagate.  The momentum lost from the jets during deceleration is transferred to the ISM, with a large portion of the momentum transfer being in the direction perpendicular to the mean jet axis.  As a result, the jets become lesser collimated, and in extreme cases the jets create a peanut-like cocoon shape, rather than well defined lobes.
\item {\bf Galaxy disc density scale-height $\zd$} - this scale height determines how fast the density decreases with distance from the Galaxy plane, such that moving further away from the Galaxy plane by $\zd$ parsec causes the density to change by a factor $e^{\rm -1}$, as described in equation \eref{rhoprof}.  Decreasing $\zd$ therefore increases the density gradient, and the density ratio between the east and west sides of SS\,433.  Increasing the density gradient means that SS\,433's west jet decelerates rapidly due to the sharp increase in density on the west side of SS\,433, in contrast to the east jet, which decelerates much less due to a rapid decrease in the ISM density over short distances to the east of SS\,433.   
\end{enumerate}

A series of jet simulations were performed to test many combinations of ($n_{\rm 0}$,\,$\zd$), in order to reproduce the correct east-west asymmetry ratio observed in W\,50.  These results are presented in \fref{jetlobes}, from which two parameter combinations appear to reproduce W50's east-west lobe extents very well, as shown in Table \tref{bestparms}: 
\begin{table}
\begin{minipage}{\columnwidth}
\centering 
\caption{Best-fit parameter values.}
\centering 
\begin{tabular}{c c c}
\hline\hline 
{\bf Parameter Set} & {$\textbf{\em n}_{\rm 0}$ (particles cm$^{-3}$)} & {$\textbf{\em Z}_{\rm d}$ (pc)} \\
\hline
 Set 1 & 0.2 & 40\\
 Set 2 & 0.1 & 30\\
\hline
\hline 
\end{tabular} 
\tlab{bestparms}
\end{minipage}
\end{table}
\medskip

\subsection{Collimation of the jet lobes}
\slab{jetcoll}
It is important to realise that the two best-fit parameter sets in \tref{bestparms} are only relevant for a jet mass-loss rate of $\dot{M}_{\rm jet}=10^{-4}{\rm M_{\odot}\,yr^{-1}}$.  This realisation comes from a particularly important result demonstrated by the dashed red-white and orange-white lines in \fref{jetlobes} corresponding to simulations {\bf \em Lobes2b} and {\bf \em Lobes3b} respectively.  These simulations show that changes in the background density normalisation $n_{\rm 0}$ and the jet mass loss rate $\dot{M}_{\rm jet}$ can become indistinguishable, as previously hypothesised in \sref{jetresults}.  For example, increasing the jet power by a factor of 10 (equivalent to increasing $\dot{M}_{\rm jet}$ if the jet speed is a constant) whilst keeping all other parameters constant, is equivalent to reducing the background density normalisation (the resistance felt by the jets) by a factor of 10.  Thus, I define the quantity $\chi_{\rm p}$ to indicate the penetrative ability of a given jet:

\begin{equation}
\chi_{\rm p} = \frac{\dot{M}_{\rm jet}}{ n_{\rm 0}}
\elab{jetpenetration}
\end{equation}
such that any two jets with the same value of $\chi_{\rm p}$ in will follow the same path in \fref{jetlobes}, for a given value of $\zd$.  Thus, for any given mass-loss rate, the corresponding best-fit value for $n_{\rm 0}$ can be calculated from \eref{jetpenetration}.

\begin{figure}
\centering
\begin{minipage}{\columnwidth}
\centering
\mygraphics{width= \columnwidth}{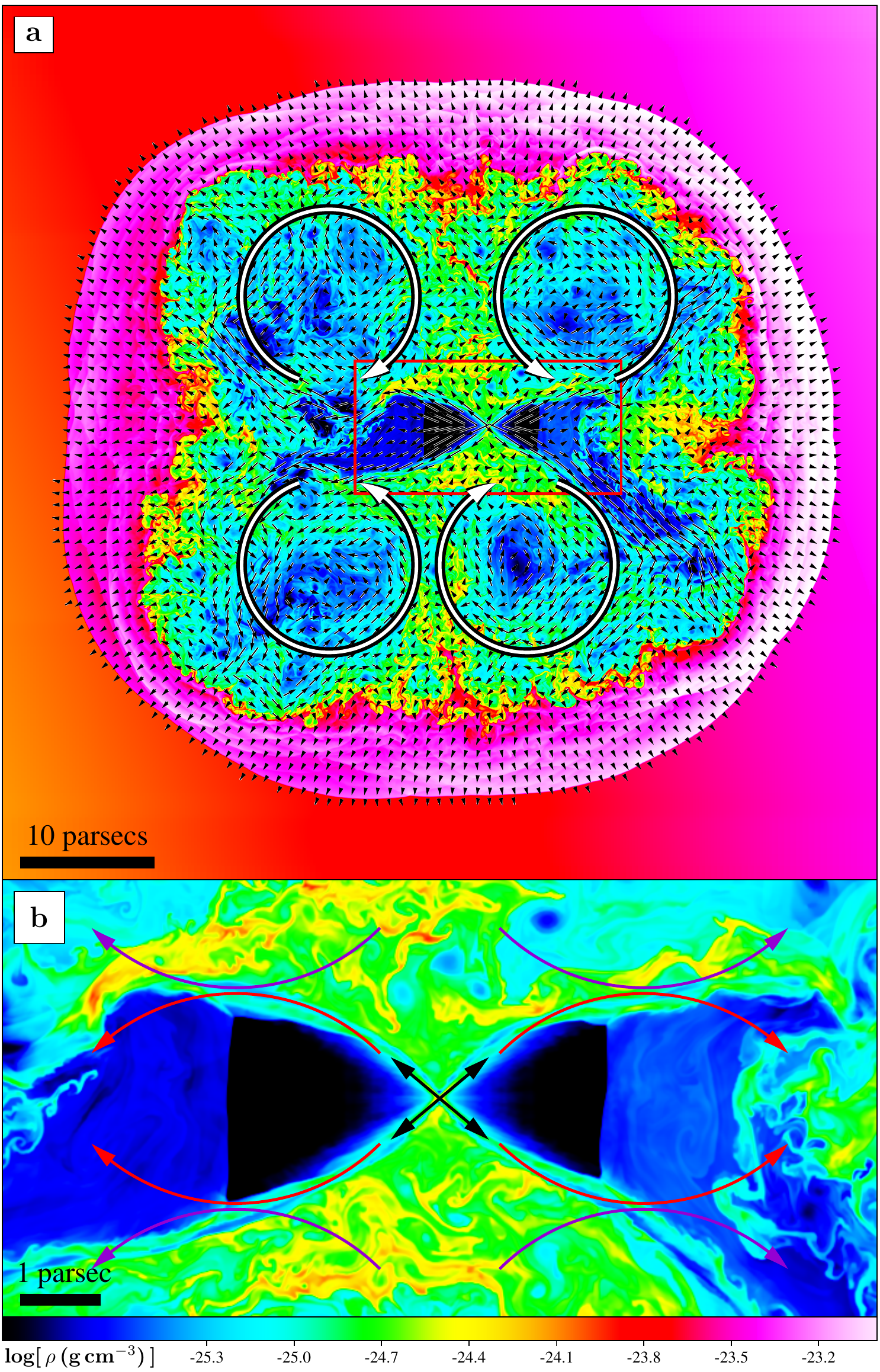}
\caption{Demonstration of the {\em kinetic} hydrodynamical refocusing of hollow conical jets.  The small black velocity vectors in (a) have magnitude proportional to the square-root of the average speed at each point, and are overlaid upon a zoomed-in density map of \J{7}.  The large white arrows in (a) show the net vortex-like movement of the turbulent gas in each quadrant.  These vortices tend to move around appreciably within each quadrant, which is the reason for the asymmetry in their locations in this snapshot.  The black arrows in (b) indicate the initial trajectory of the jets.  The purple arrows indicate the turbulent motion of the agitated gas within the jet cocoon, and the red arrows indicate the subsequent interaction and refocusing of the jets back towards the jet-axis.  A short animation showing this effect is avilable at: \hbox{\phref{blue}{http://www-astro.physics.ox.ac.uk/~ptg/RESEARCH/research.html}{\small $http://www$-$astro.physics.ox.ac.uk/$$\sim$$ptg/RESEARCH/research.html$}}}
\flab{refocus}
\end{minipage}
\end{figure}

\subsection{Hydrodynamic refocusing of SS\,433's jets.}
\slab{refocus}
It has been suggested that astrophysical jets of the hollow conical type, may undergo refocusing towards the centre of the cone \citep{eichler1983} due to the ambient pressure from the ISM through which the jet propagates.  However, we find evidence of refocusing of our hollow conical jets through a different mechanism, which is purely dependent upon the kinematics of the ISM within the cocoon of the jet. Two subtly different refocusing mechanisms are suggested here, but both are based upon momentum exchange between the jet and its environment, as follows:

\begin{enumerate}
\item {\bf Kinetic refocusing}~- the jets are hydrodynamically refocused via collisions with the low-density but fast-moving and turbulent ISM within the jet cocoon, which has been energised through past interactions with earlier jet ejecta.  Due to the low density of the gas within the jet cavity, it is quite easy for this material to be accelerated to high speeds through interactions with the jets.  After gaining momentum from the jets, the fast-moving gas interacts with the wall of the cocoon and develops a vortex-like rotation, as indicated by the arrows of the velocity field of the gas superimposed onto \fref{refocus}a.  In this case, momentum transfer from the jets to the cocoon ISM has the dual effect of altering the jet trajectory and further stirring the turbulent ISM motion.  This effect is self-sustaining once it begins.  This effect is seen in all of our simulations that have a constant jet cone angle greater than zero, such as \J{5}, \J{6} and \J{7} of Figures \ref{fig:jetfig}e, \ref{fig:jetfig}f and \ref{fig:jetfig}g.  The process of momentum exchange with the turbulent ISM is depicted in \fref{refocus}b.
\item {\bf Static refocusing}~- a more gradual focusing mechanism occurs when the jets interact at a grazing angle with the dense, but comparatively slow-moving, cavity walls of the jet cocoon.  The interaction occurs at a grazing angle when the jet cocoon is sufficiently elongated, such as that created by a cylindrical jet (or a jet with a near-zero cone angle).  Due to the high density of the cavity wall, it experiences a much smaller acceleration per unit momentum-exchange than does the low-density gas of the kinematic refocusing method.  As a result, the jet is redirected back towards the mean jet axis upon collision with the cavity walls.  This mechanism is seen to occur in \J{8} of \fref{jetfig}h. 
  \end{enumerate}


\begin{table*}
\begin{minipage}{17cm}
\centering
\caption[Summary of SNR-jet simulation parameters]{A description of the key parameters relevant to each of the SNRJet simulations.}
\begin{tabular}{c c c l}
\hline\hline 
{\bf Shortname} & {\bf Description} & \multicolumn{2}{c}{{\bf Parameter Details}} \\
\hline
\SNR{A} & SNR only &  \multicolumn{2}{c}{SNR defaults and:~$\rho$-profile\,=\,1, v-prof\,=\,1, $n_{\rm 0} = 0.1\,{\rm cm^{-3}}$, $\zd=40$\,pc} \\
\SNR{B} & SNR only &  \multicolumn{2}{c}{SNR defaults and:~$\rho$-profile\,=\,1, v-prof\,=\,1, $n_{\rm 0} = 0.2\,{\rm cm^{-3}}$, $\zd=30$\,pc} \\
\hline
SNRJet1 & \SNR{A} with a modified \J{1} &  \multicolumn{2}{c}{Modified parameters:~  $n_{\rm 0} = 0.1{\rm cm^{-3}}$ } \\
SNRJet2 & \SNR{B} with a modified \J{1} &  \multicolumn{2}{c}{Modified parameters:~  $\zd=30$\,pc}\\
SNRJet3 & \SNR{A} with a modified \J{6} &  \multicolumn{2}{c}{Modified parameters:~  $\theta_{\rm 0}=20^{\circ}$}\\
SNRJet4 & \SNR{A} with a modified \J{8} &  \multicolumn{2}{c}{Modified parameters:~  $\dot{\theta}=20^{\circ}/1500\,yr$} \\
SNRJet5 & \SNR{A} with a modified \J{9} &  \multicolumn{2}{c}{Modified parameters:~  $\theta_{\rm 1}=0^{\circ}$, $\theta_{\rm 2}=20^{\circ}$} \\
\hline
\multicolumn{4}{c}{Jet default parameters: $\dot{M}_{\rm jet}=10^{-4}\,{\rm M}_{\odot}\,{\rm yr^{\rm -1}}$} \\
\hline
\hline 
\end{tabular}
\tlab{SNRjetsims}
\end{minipage}
\end{table*}
\medskip

\section{SNR + Jets:  Results and Discussion}
\slab{snrjets}

Following from Table \tref{bestparms} in \sref{jetlobes}, the combination of \hbox{parameters} $(n_{\rm 0}=0.2\,{\rm cm}^{-3},\zd=40\,{\rm pc})$ and $(n_{\rm 0}=0.1\,{\rm cm}^{-3},\zd=30\,{\rm pc})$ were determined to be most representative of the observations of W\,50, for a jet mass loss rate of $\dot{M}_{\rm jet}=10^{-4}{\rm M_{\odot}\,yr^{-1}}$.  Although both of these parameter combinations are equally feasible, they are also almost equivalent in terms of the jet morphology produced.  Hence, we focus on the first parameter set \hbox{$(n_{\rm 0}=0.2\,{\rm cm}^{-3},\zd=40\,{\rm pc})$} as being the best description of SS\,433's environment, and the effects of the interaction between the jets and SNR are investigated for this setting.

All jet models featured here use the same jet speed of 0.26c corresponding to the observed speed for SS\,433's current jet state.  We acknowledge that the jet speed is one of the variable parameters that can change between outbursts of jet activity, and the choice to keep the jet speed constant here has no physical basis other than to limit the number of free parameters in the models.  This enables us to focus on the effects of changing the jet precession cone angle with each different jet outburst.  The effects of varying other parameters such as jet speed and jet mass loss rate, will be investigated in a subsequent publication.

\subsection{Simulating the Jet-SNR interaction}
\slab{snrjetinteraction}

The best-fit parameter sets from Table \tref{bestparms} were used to create two new SNR simulations: \SNR{A} and \SNR{B}, thereby incorporating the most appropriate background environment settings for the evolution of the new SNR shells.  These new SNR simulations are otherwise identical to \SNR{6}.  As per \sref{presedov}, the SNR evolution is followed until the radius approaches 45 parsec, which happens at $\approx\,$21,000\,yrs for \SNR{A}, and $\approx\,$17,000\,yrs for \SNR{B}, due to the slightly lower density of \SNR{B}.  Various jet models were then invoked to investigate the SNR-jet interaction, and parameters tested in each simulation are described in Table \tref{SNRjetsims}.

\subsection{Describing each SNR-Jet interaction}
\slab{snrjetresults}
The final snapshot for each of the simulations described in Table \tref{SNRjetsims} are shown in \fref{snrjetfig}, where the left column shows the density maps, and the right column shows the gas speed $(v_{\rm x}^2+v_{y}^2)^{\rm \frac{1}{2}}$.  The purple contours on the left column of \fref{snrjetfig} show the outline of W\,50 from the VLA radio observations of \citet{Dubner}.  The green contours on the right column of \fref{snrjetfig} show the structure of SS\,433's jet lobes from the ROSAT X-ray observations of \citet{brinkmann1996}.  The inlay on each panel shows a 4$\times$ magnified region about the coordinates of the jet launch.

\subsection{The roles of ambient density \textbf{n$_{\rm 0}$} and scale-height \textbf{Z$_{\rm d}$}}
\slab{snrjet1}
\SJ{1} of \fref{snrjetfig}(a \& b) shows how \J{1} from \fref{jetfig}a would interact with \SNR{A}.  The extents of the east and west jet lobes are in good agreement with the radio observations of W\,50 as shown by the purple contours in \fref{snrjetfig}(a).  This confirms that the parameters were determined correctly \hbox{$(n_{\rm 0}=0.2\,{\rm cm}^{-3},\zd=40\,{\rm pc})$}, and that the presence of the SNR has little effect upon the east-west lobe asymmetry along the jet-axis direction.  The SNR does however widen the jet at the point where the jet penetrates the SNR shell, in a similar manner to waves diffracting through an aperture.  This is because the jet breaks through the relatively high density of the compressed SNR shell into the surrounding lower-density ISM, and the impact of the jets with the shell increases the momentum of the SNRJet material in the direction perpendicular to the jet axis.  Note that the jet lobes produced in \SJ{1} have a higher degree of collimation than those of \J{1} from \fref{jetfig}a, due to the background density normalisation being a factor of 5 lower in \SJ{1} than \J{1}.  Judging by the purple contours, the east lobe of \SJ{1} seems to be approximately equal in width to the eastern radio lobe, whereas the transition between the SNR shell and the western lobe seems to be more gradual in the purple radio contours than the simulation would indicate.  This could be due to a previous jet episode impacting on the younger SNR at an earlier time, such that the SNR shell is inflated more rapidly at the location of the jet impact.   The green contours in \fref{snrjetfig}(b) indicate again that the eastern jet lobe has similar dimensions for both the simulation and X-ray observations, however the western lobe X-ray emission is abruptly truncated short of the full extent of the simulated jet lobe.  The X-ray lobe is both wider and shorter than the west lobe of the simulation, as though the X-ray emitting jet material is hitting a dense wall of material, which is true if the increased radio brightness of the nebula at this position indicates an increase in density (see point D in \fref{RX}).  Again, this could be indicative of a previous jet ejection episode.

\SJ{2} of \fref{snrjetfig}(c \& d)  is similar to \SJ{1} in most respects, except for the fact that the jet lobes are more highly collimated than \SJ{1} due to a decrease in the density normalisation by a factor of 2.  The jet lobes produced by \SJ{2} are thinner than the radio/X-ray lobes as indicated by the purple/green contours, and this probably means that the density of $n_0 = 0.1$ particles cm$^{-3}$ is too low, and it's likely that \SJ{1} with $n_0 = 0.2$ particles cm$^{-3}$ is more representative of the environment at the locality of SS\,433.

\begin{figure*}
\centering
\vspace{-0.2cm}
\begin{minipage}{\textwidth}
\centering
\mygraphics{width= \textwidth}{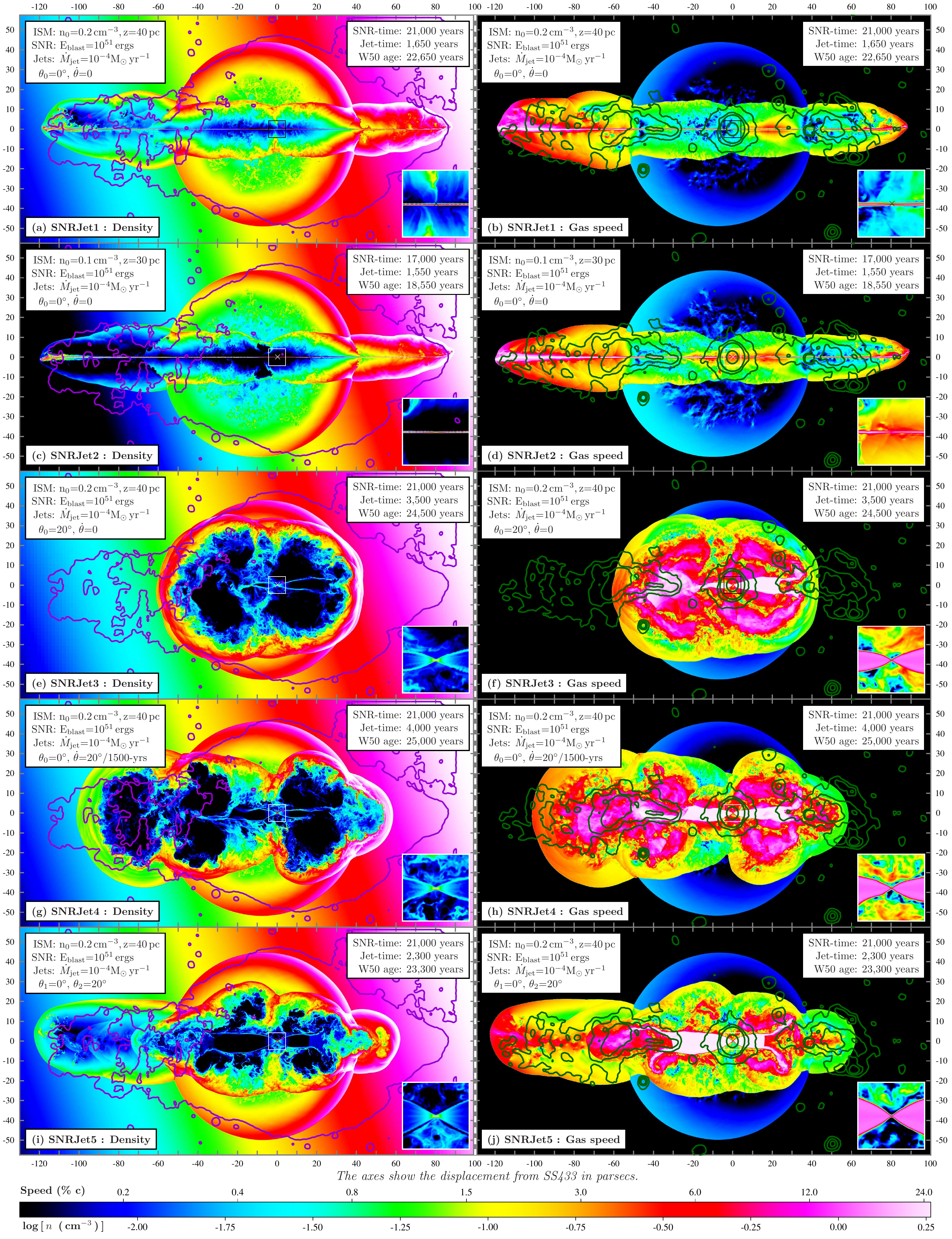}
\caption[Interaction of jets with an evolved SNR]{\footnotesize This figure illustrates the varying nebular morphologies produces through the interaction of 5 different jet models with an evolved SNR.  Refer to Table \tref{SNRjetsims} for a summary of the differences between these simulations, and \sref{snrjets} for a detailed commentary on the interpretation of each one.  The animations associated with these simulations are available at: \hbox{\phref{blue}{http://www-astro.physics.ox.ac.uk/~ptg/RESEARCH/research.html}{\small $http://www$-$astro.physics.ox.ac.uk/$$\sim$$ptg/RESEARCH/research.html$}}.}
\flab{snrjetfig}
\end{minipage}
\end{figure*}

\subsection{The role of precession cone opening angle}
\slab{snrjet3}
\SJ{3} of \fref{snrjetfig}(e \& f) features a precessing jet with half-cone angle of 20$^{\circ}$, very similar to that of the current jet in SS\,433.  Although this jet model doesn't reproduce the lobes of W\,50, it does accelerate the expansion of the east and west hemispheres of the SNR bubble.  This is further evidence that a previous jet episode with a large cone angle probably contributed to the shape of W\,50.  It is unlikely that this inflation of the east and west hemispheres were caused by the current jet episode in SS\,433 due to time arguments, as explained in \sref{snrjet5}.  This hollow conical jet model is also affected by the {\em Kinetic Refocusing} mechanism described in \sref{refocus}, and the effects are noticeable at around 500 years onwards in the simulation (refer to the simulation movies at the accompanying weblink).

\subsection{The role of time-varying precession cone opening angle}
\slab{snrjet4}
\SJ{4} of \fref{snrjetfig}(g \& h) features a conical jet whereby the jet precession cone half-angle linearly increases from 0$^{\circ}$ to 20$^{\circ}$ over a period of 1500 years.  The jet parameters are almost identical to that of \J{8} from \sref{jetresults}, with the only difference being the time taken for the jet to reach the maximum cone angle of $\sim$20$^{\circ}$, which was 1000 years rather than 1500 years.  However, the difference in the nebular morphology produced by \J{8} and \SJ{4} is striking (compare \fref{jetfig}h with \fref{snrjetfig}g), and this difference is due entirely to the presence of the compressed SNR shell.

The hypersonic shockwave created by the supernova explosion sweeps up most of the gas in the ISM, and the density distribution from the final snapshot of \SNR{A} is shown in \fref{snrdensityprofile}, along with the original density profile of the background ISM before the supernova explosion.  The morphological difference between \J{8} and \SJ{4} demonstrates (yet again) the importance of the ISM density distribution in the vicinity of the jet launch site, and the profound influence this can have upon the resulting jet cocoon.

\begin{figure}
\centering
\begin{minipage}{\columnwidth}
\centering
\mygraphics{width=\columnwidth}{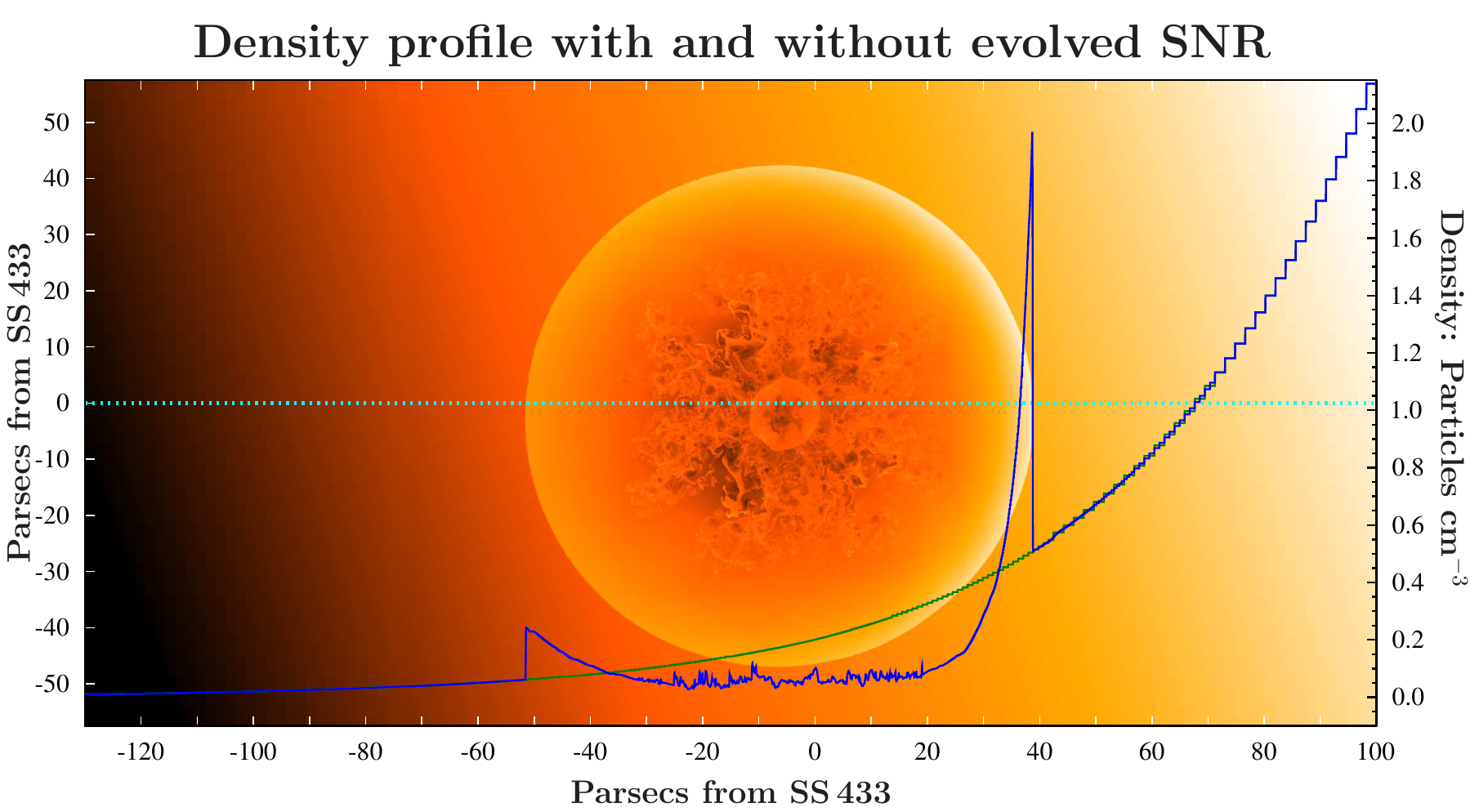}
\caption{The image shows a density map (logarithmic colour scale, white\,=\,high density, black\,=\,low density) of \SNR{A} in the final stages of evolution. The blue line shows a density line profile taken through the centre of the displayed image (indicated by the dotted line).  The green line shows the same density line profile of the unperturbed ISM (before the supernova explosion occurred).  This figure illustrates the difference in the local density distribution as seen by \J{8} (green line, without SNR), and \SJ{4} (blue line, after SNR has ploughed through the ISM).  The lower density environment through which the jet from \SJ{4} propagates, is the reason for the morphological differences with \J{8}.}
\flab{snrdensityprofile}
\end{minipage}
\end{figure}

\subsection{The role of episodic jet ejection}
\slab{snrjet5}
\SJ{5} of \fref{snrjetfig}(i \& j) uses an episodic jet model with two outbursts; the first jet outburst features a cylindrical jet which lasts for 200 years, and the second outburst is a precessing conical jet with half-angle 20$^{\circ}$ of 2300 years in duration.  \SJ{5} displays both {\em Kinetic} and {\em Static refocusing} (see the movies on the accompanying webpage).  \hbox{\em Kinetic refocusing} \hbox{becomes} \hbox{noticeable} around $\sim$700 years after the jet is first switched on, and is readily recognisable as four regions of swirling gas within the SNR, which are (approximately) symmetrically located about the jet launch point (one in each quadrant).  {\em Static refocusing} begins approximately 2000 years after the jets are switched on, whereby the jets are refocused towards the jet axis upon interaction of the denser jet cocoon gas located near the SNR shell, which was penetrated several hundred years earlier by the cylindrical jet. 

This jet model was devised to combine the successes of both cylindrical jets \SJ{1} in reproducing the correct jet lobe extents, and the conical jets \SJ{3} in producing the required inflation of the east and west sides of the SNR such that a gradual transition from SNR shell to lobes is created.  However this attempt was unsuccessful, because the cylindrical jets (despite the short episode duration of just 200 years) exceed the extent of W\,50's lobes in the time it takes for the conical jets to catch up with the SNR shell and contribute to its expansion, as shown in \fref{snrjetfig}i.

\subsection{Jet Kinematics}
\slab{jetkin}
Astrophysical jets sweep up the ISM just as supernova blast waves do, with the very important difference that supernova explosions are very short-lived (in astrophysical terms), injecting enormous amounts of energy into the ISM in one event.  For jets, the ejection of high-speed material often continues for long periods of time, and in fact many thousands of years of jet activity are required to create the W\,50 nebula surrounding SS\,433.  As a consequence, the properties of the environment surrounding the jet-launch site, change with time.  The activity of the jets clears a path through the ISM for the jet material that is ejected at later times.  This is one possible explanation for the variable observational characteristics of microquasar jets seen at different times or outbursts, as mentioned in \sref{outbursts}.  However, a more obvious result from these simulations is that the jet suffers very little deceleration inside the jet cocoon or cavity produced by earlier jet activity, and most of the deceleration occurs in the final stages when the jet ejecta catches up with the dense build-up of gas behind the jet shock front.  This is illustrated in the right-hand column of \fref{snrjetfig} where the colour maps show the gas speed; the pink/white regions show gas travelling very close to the launch-speed.  The full details of this effect are given in a companion paper (how to site this companion paper?).

\subsection{Jet symmetry and collimation}
\slab{collimation}

The level of collimation is actually higher for pulsed jets than for continuous jets.  For a continuous jet, although the jets receive constant replenishment of the kinetic energy and therefore don't show signs of deceleration, a side-effect is that the jet bunches-up and crashes into the jet in front.  Any tiny asymmetric displacement of the bunched-up jet about the jet axis will cause the upstream jet to transfer an increased amount of momentum in the transverse direction.  An analogy for this effect is drawn with that of a cue-ball struck off-centre by the cue.  This effect is much less noticeable when the jet is pulsed, presumably due to a lower occurrence of jet blockages.

\subsection{Limitations of current physical models}
\slab{disclaimer}
\corx{Due to the microphysics not being well understood, comparison between observation and simulations are not straightforward.  Since we are concerned with the general morphology of the W\,50 nebula, only the overall brightness profile from the radio observations should be compared with \fref{jetfig} and \fref{snrjetfig}.  Optical observations of W\,50 \citep{Boumis2008} show small-scale filamentary structures; simulations of these filaments would require a detailed description of the relevant microphysics (such as the amount of electron heating in collisionless shocks, the detailed structure of SN ejecta).  We note that the presence of any clumpyness of the circumstellar medium and possibly dust would have an effect on the propagation of the young SNR and jets.  The density maps from our simulations reproduce the key morphological aspects of W\,50 such as the lobes with appropriate lengths and widths, and a circular region away from the jets due to the expansion of the SNR, in good agreement with the radio observations.  The X-ray observations appear to trace out regions of recent jet activity, but again only the general large-scale shape is important for comparison with our simulations.}

\begin{figure}
\centering
\begin{minipage}{\columnwidth}
\centering
\mygraphics{width=\columnwidth}{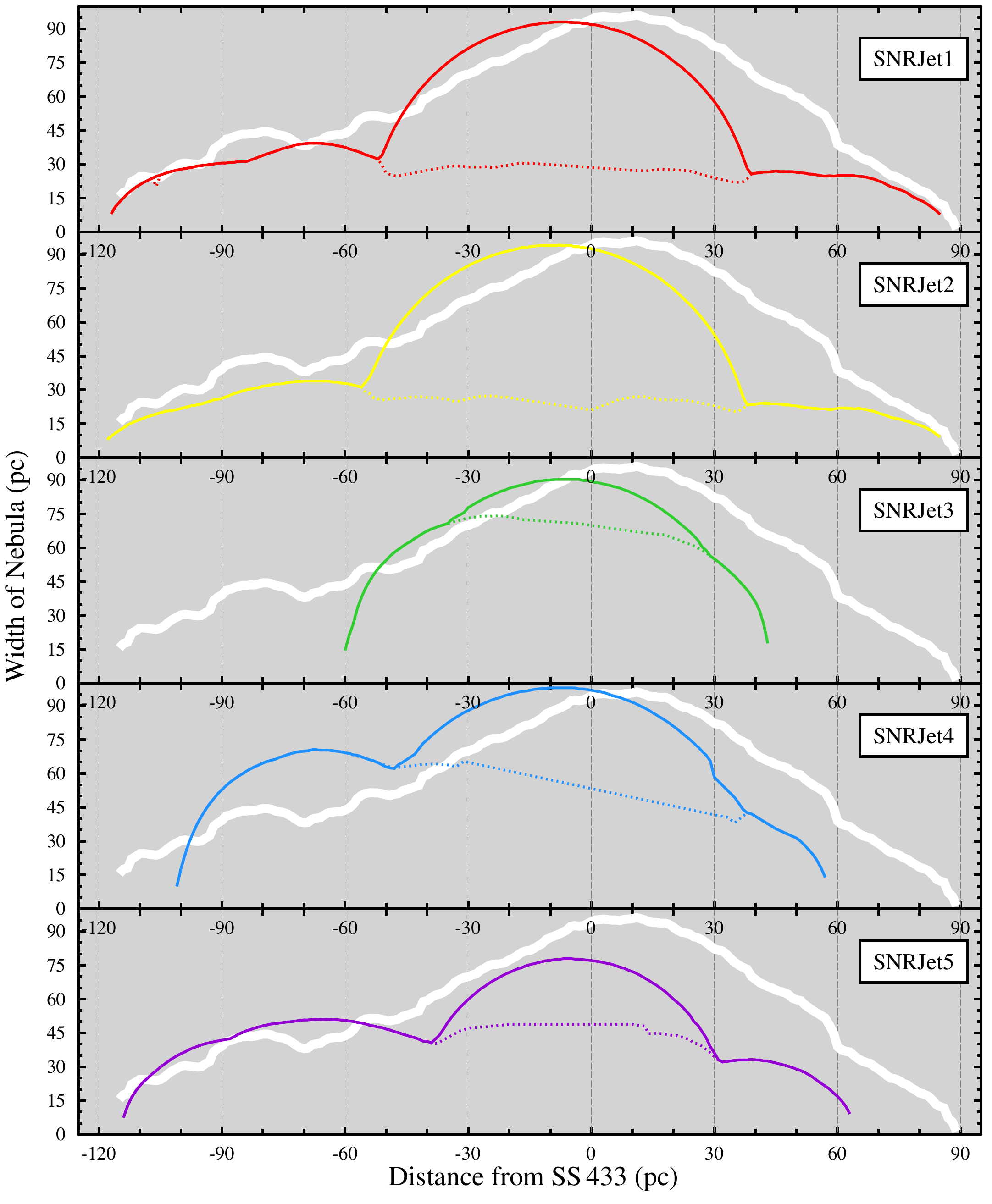}
\caption{\small The vertical width of each nebula produced from the SNR-jet interaction simulations from \fref{snrjetfig}, is shown as a function of distance from SS\,433 by the coloured lines in each panel.  The width of W\,50 as a function of distance form SS\,433 is also shown by the thick white line in each panel, as measured from the \citet{Dubner} image.  Note how the transition between SNR and jet lobes is very smooth in reality (white lines), and in fact it is difficult to say where the SNR shell ends and the jet lobes begin when viewed in this way.  The dashed coloured lines show the width of the jet internal to the SNR.  The bumps in the lobes indicate the locations of the annular features in W\,50's lobes. }
\flab{lobewidth}
\end{minipage}
\end{figure}

\section{Conclusions}
\slab{conclusions}
The cylindrical jet models reproduce the lobes of W\,50 very well in terms of their absolute and relative extents in the east and west directions.  As \fref{snrjetfig} and \fref{lobewidth} show, the cylindrical jets create jet lobes with well-defined boundaries at the site of the impact between jets and the SNR shell.  This is not in agreement with the radio observations of \citet{Dubner} in which there is a more gradual transition between the circularity of the SNR shell and the elongated lobes (see \fref{lobewidth}).  

The conical jet models do produce lobes with a lower degree of collimation appropriate to W\,50's lobes very near to the SNR shell, however these jets continue to diverge away from the jet axis and prove too wide at larger distances from SS\,433 to be considered an accurate representation of W\,50's lobes.  

It was hypothesised that the smooth transition between SNR shell and lobes might be created if a conical jet model is used in conjunction with a cylindrical jet model.  This was tested in \SJ{5} where a cylindrical jet (an earlier jet episode) was invoked before a conical jet with the precession cone half-angle approximately equal to that currently observed in SS\,433.  However, this leads to problems due to the large time interval (at least 2000 years) required for the conical jets to reach the surface of the SNR shell, during which time the cylindrical jet travels beyond the boundaries of the simulation domain, indicating that a real jet would have far exceeded the extent of W\,50's lobes in the time required by the conical jets.  This problem is eradicated if the conical jets begin first and are allowed to evolve for as long as is necessary before the cylindrical jets begin, because the latter only require of order $\sim$1600 years to reach the required lobe extent.  However the situation gets more complex in that the final jet episode must be in agreement with current observations, and so a third jet episode representing SS\,433's current jet precession state must be invoked in order to satisfy this requirement.  The role of episodic jet outbursts will be further investigated in future work. 

In summary, it seems clear that neither cylindrical nor conical jets can independently reproduce the interesting morphology displayed by W\,50, and it seems several jet episodes from SS\,433 with varying jet characteristics are needed in order to sculpt W\,50 in the ISM.

\subsection{Energetics of the SS\,433-W\,50 system}
\slab{energetics}
With a radius of 45\,pc, W\,50 is among the largest SNR observed to date, and to consider higher SNR blast energies of 10$^{52}$ ergs or so, would be reasonable.  This would usefully lessen the latency period between the supernova event and the jets switching on, since the SNR expansion speed begins as $\sqrt{2\,E_{\rm blast}/M_{\rm ej}}$.  Such high blast energies would be expected for a very massive progenitor.  

If both stars in the binary formed at the same time, a simple main-sequence lifetime argument requires that the SS\,433's progenitor must have been more massive than its Wolf-Rayet companion star at 24M$_{\odot}$ \citep{Blundell4}, since it was first to detonate.  A massive Wolf-Rayet progenitor producing a Hypernova explosion includes the possibility that SS\,433 may have formed through a Gamma Ray Burst (GRB) event.  To simulate this scenario, our models would need to be adapted to include a short but highly relativistic jet outburst simultaneous with the progenitor detonation, to be investigated in a further publication.

The rapidly expanding SNR could make W\,50 an ideal candidate for producing of high energy cosmic rays.  We note that with the large radius of 45\,pc for W50 and the Galactic field strength of 5\,$\mu$Gauss makes confinement of particles with energies up to 2$\times$10$^{\rm 17}$eV possible.  

\subsection{The Radio and X-ray correlation}
\slab{RX}
Our simulation results can be shown to be consistent with radio and X-ray observations of W\,50 as shown in \fref{RX}, in conjunction with the jet outbursts hypothesis.  Note the striking correspondence in \fref{RX}(b) between the low-level (blue) emission and the outline of the contour from the radio nebula.  Furthermore, the bright filament at point {\bf A} on the far west of W\,50 is coincident in both X-rays and radio.  It is equally interesting to note that the brightest parts of the X-ray lobes (emission from the jets) at points {\bf B} and {\bf C} are almost exactly coincident with regions where the radio emission is faintest (i.e. radio-quiet regions where  the X-ray lobes are brightest).  Finally, the point labeled {\bf D} shows an abrupt termination of the eastern X-ray jet lobe followed by an abrupt brightening of the radio jet lobe.  This could indicate two different populations of particles, perhaps where the relatively recent, hot, and X-ray emitting jet ejecta from a new jet episode has caught up with the older, cooler, ejecta from an outburst that happened much further in the past.

It is generally noted that the X-ray emission from W\,50 is seen to be confined to a region much closer to the mean jet axis than the precession cone of SS\,433's jets would indicate, and this could be attributed either to a previous jet ejection episode with a precession cone angle close to zero, or to the mechanism of refocusing of SS\,433's jets as detailed in \sref{refocus}. 
\begin{figure}
\centering
\begin{minipage}{\columnwidth}
\centering
\mygraphics{width=\columnwidth}{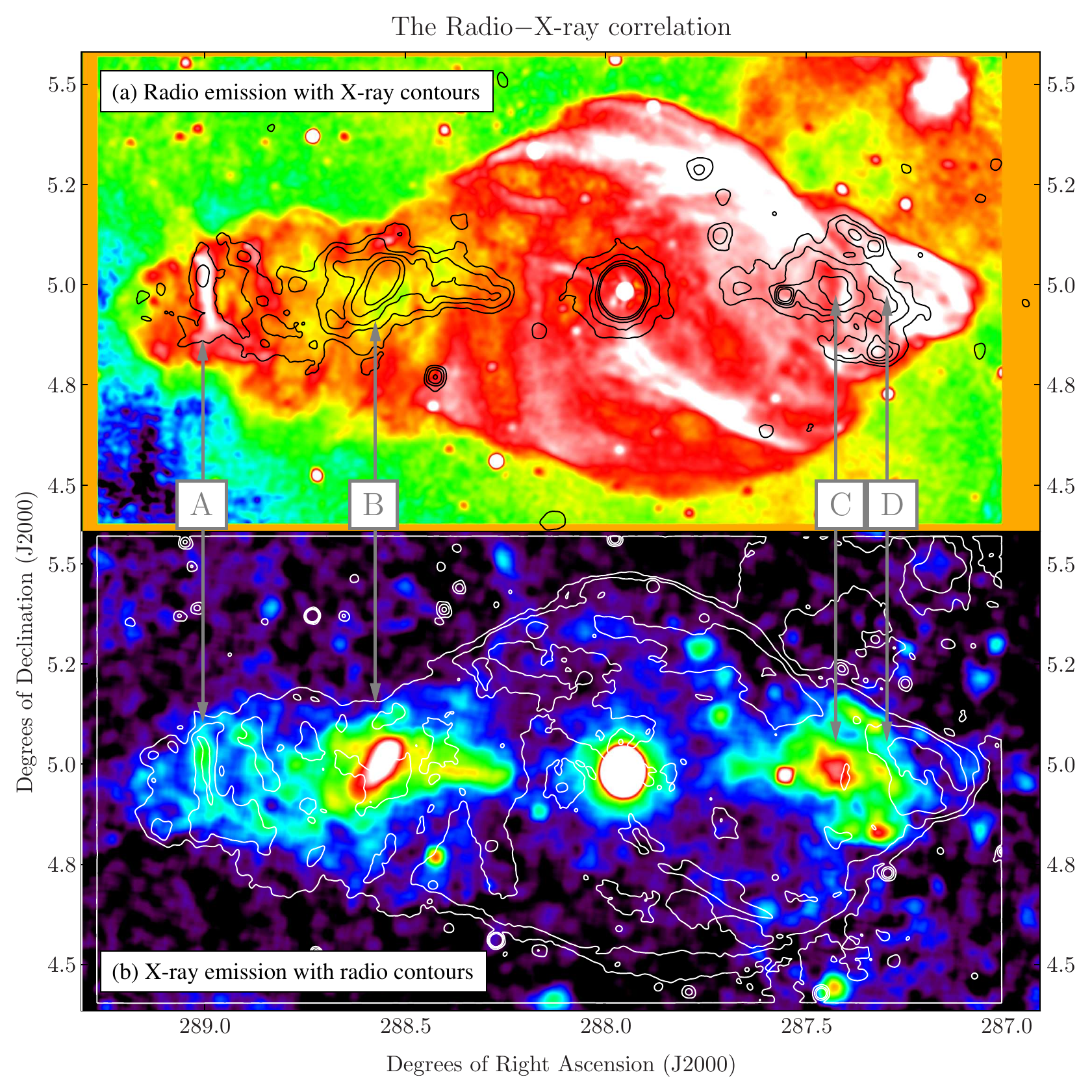}
\caption{The striking similarities and differences between (a) the VLA radio image of W\,50 \citep{Dubner} and (b) the ROSAT X-ray image \citep{brinkmann1996} are entirely consistent with the hypothesis of episodic jet ejection in SS\,433.  The ROSAT contours are overlaid upon the radio image in (a), and the radio contours are overlaid upon the ROSAT image in (b).}
\flab{RX}
\end{minipage}
\end{figure}

\subsection{Comparison with previous work}
\slab{grandslam}
Despite the appeal of some of the previous hydrodynamic models attempting to reproduce W\,50, there are some inconsistencies which must be discussed in order to begin to fully understand the nature of the SNR-jet interaction in SS\,433 and W\,50.   
First, we note that concrete graphical evidence in the literature supporting claims of having reproduced the annular structure in the lobes of W\,50 through hydrodynamic simulations, is (so far) scant.  In Fig. 3 of \citet{velazquez2000}, the pressure contour at 1150 years shows the emergence of two annuli through the jet-SNR interaction, however it is not clear if these are artefacts of the reflective boundary conditions imposed.  The computed synchrotron emission (Fig. 5 of \citet{velazquez2000}) does resemble the annular structure in the eastern lobe of W\,50 at approximately 3000 years after the jets are initiated.  However, it is easy to create artificial annular structures through azimuthal revolutions with axial symmetry.  The follow-up paper \citet{zavala2008} with improved jet models and in three dimensions, reports no evidence of any annular structure.  They acknowledge that the annular structures created in the previous paper are a possible consequence of the reflective boundary conditions imposed therein, and the undesirable effects produced via axisymmetric codes with reflective boundary conditions are well known to those authors \citep{raga2007}.

Each of the simulations shown so far in the literature have featured scaled-down versions of the W\,50 system.  The primary assumption for any work involving the SS\,433-W\,50 system, is that both objects are at the same location, and independent methods have determined a distance of 5.5\,kpc is appropriate to each of SS\,433 \citep{Blundell2} and W\,50 \citep{Lockman}.  At this distance, the radius of the SNR is 45 parsecs and the jet extents are 86.5 and 121.5 parsecs for the west and east jet lobes respectively, as indicated in \fref{allscales}.  These physical sizes (and their ratios) are of paramount importance in forming a consistent model of the SS\,433-W\,50 interaction, that is in agreement with observations.  We stress that (in contrast to the claims of \citet{zavala2008}), it is {\em not} sufficient to simply scale-up simulated models of the system for comparison with observations.  The reasons for this are as follows:
\begin{enumerate}
\item The absolute extents of the jet lobes at a given time are characteristic of the integrated column density along the path of the jets (a measure of the resistance faced by the jets) as they travel through the ISM.  Similarly, the radius of the SNR at a given time is characteristic of the mass it has swept up, which depends roughly on the cube of the radius.  Aside from the density normalisation, the evolution of each of these is also dependent upon the energy input, as governed by $E_{\rm blast}$ for the supernova, and $\dot{M}_{\rm jet}$ over some time period for the jets.  The jets propagate much faster than the SNR at all times, and so a morphological match to W\,50 at some scaled-down radius of the SNR will be lost at later times.  In essence, the jets would extend far beyond the domain of W\,50 by the time the SNR has expanded to a radius that is in agreement with observations.
\item The relative extents of the east and west jet lobes are characteristic of the exponential density background, and the ratio of the lobe extents is not linearly scalable with either total jet travel time or total lobe extent, as shown in  \fref{jetlobes}. 
\item Simulating the absolute extents of both SNR and jet lobes is important as this allows us to place constraints on the formation time and energetics for each component, and has relevance to the aforementioned latency period. 
\item Jets could possibly inflate (accelerate the shell expansion of) a small SNR corresponding to some scaled-down version of W\,50.  However, we find that this is not the case for a SNR as large as 45\,pc, and the jets have absolutely no effect upon the SNR expansion in the direction perpendicular to jet launch.
\end{enumerate}

\subsection{Summary}
\slab{summary}
Of all simulations presented in this paper, only cylindrical jets have produced sufficiently collimated jet lobes comparable to the jet lobes in W\,50.  Simulations mimicking the jet precession cone observed in SS\,433 at the current epoch, do not reproduce the lobes of W\,50.  Neither cylindrical, nor conical jets resembling SS\,433's current precession state, are able to reproduce the interconnecting region between W\,50's circular SNR shell and the jet lobes, although very large cone angles of 40$^{\circ}$ can produce it.  Our findings therefore suggest that at least three very different jet states are required during SS\,433's very active jet history, in order to produce (i) well collimated jet lobes, (ii) smoothly connects SNR and lobes, (iii) the precessing corkscrew jet observed at the present day.  This is the minimum number of jet outbursts required to create a W\,50 shaped nebula around the microquasar, but it is possible than many more jet outbursts have made contributions over time.  Further evidence of episodic jet outbursts from SS\,433 comes from the comparison of the radio and X-ray observations detailed in \sref{RX}, and from an archival radio study of W\,50's kinematics described in a companion paper (cite the companion paper), the results of which are entirely consistent with the hydrodynamic simulations presented here.

The resultant nebula morphology created through sequential episodic jet activity is {\em strongly} dependant upon the nature of the jet activity that occurred previously to it.  Early jet activity ploughs gas away from the point of jet-launch, thereby creating a cavity of lower ISM density through which later jet ejecta can travel more easily.  Nearby ISM that has been agitated or stirred by the influence of the jets, can create vortices and bulk flows of gas which kinetically refocus conical jets, through exchange of momentum.  This jet refocusing also occurs when the jets interact with the dense cavity walls behind the jet-lobe shock front.

The focal length associated with the kinematic refocusing mechanism of \sref{refocus} increases slightly with the precession cone angle and the size of the jet cavity, but is typically of order 10\,pc or so.  The focal length associated with the static refocusing mechanism of \sref{refocus} also depends upon the precession cone angle, and the width of the jet cavity carved out by the earlier jet ejecta, but at 30$-$40\,pc it is typically much larger than the focusing length induced by the kinematic mechanism.  As such, the focusing length for SS\,433's jets should tell us something about the history of its jet activity, and an investigation is under-way to observe this.  We stress however, that while these are both interesting hydrodynamical effects, neither method of jet refocusing produces the level of collimation required to describe the lobes of W\,50, based on SS\,433's current jet precession state.    

We further conclude that careful hydrodynamical simulations can provide useful constraints upon the parameters of complex astrophysical systems.  The distinct morphology of the W\,50 complex has allowed us so determine the following parameters as being most appropriate in describing SS\,433's environment: ($n_{\rm 0}=0.2$cm$^{-3}$, $z=40$pc) and ($n_{\rm 0}=0.1$cm$^{-3}$, $z=30$pc).

\section*{Acknowledgments}

We would like to express our gratitude to Jocelyn Bell Burnell and Sebasti\'an P\'erez for some very interesting discussions and for suggesting useful improvements to this manuscript. We warmly thank James Binney, Philipp Podsiadlowski and Katrien Steenbrugge for some very interesting discussions.  We are very grateful to Jonathan Patterson for providing top-notch computer support and round-the-clock maintenance of the Glamdring cluster.  P. T. G. would like to thank the Science and Technology Facilities Council for a D.Phil Studentship.  K. M. B. thanks the Royal Society for a University Research Fellowship.  The software used in this work was in part developed by the DOE-supported ASC / Alliance Center for Astrophysical Thermonuclear Flashes at the University of Chicago.  Portions of the analysis presented here made use of the Perl Data Language (PDL) developed by K. Glazebrook, J. Brinchmann, J. Cerney, C. DeForest, D. Hunt, T. Jenness, T. Luka, R. Schwebel, and C. Soeller and can be obtained from http://pdl.perl.org. PDL provides a high-level numerical functionality for the Perl scripting language \citep{glazebrook97}.

\bibliographystyle{mn2e} 
\bibliography{W50_MNRAS} 

\end{document}